# Post-Quantum Cryptography for Intelligent Transportation Systems: An Implementation-Focused Review

**Abdullah Al Mamun [1,*], Akid Abrar [2], Mizanur Rahman [2], M Sabbir Salek [3], and Mashrur Chowdhury [1]**

[1] Glenn Department of Civil Engineering, Clemson University, Clemson, SC 29634 USA. E-mail addresses: abdullm@clemson.edu (A.A. Mamun); mac@clemson.edu (M. Chowdhury).

[2] Department of Civil, Construction, and Environmental Engineering, The University of Alabama, Tuscaloosa, AL 35487 USA. E-mail addresses: aabrar@crimson.ua.edu (A. Abrar); mizan.rahman@ua.edu (M. Rahman).

[3] National Center for Transportation Cybersecurity and Resiliency (TraCR), Greenville, SC 29607 USA. E-mail address: msalek@clemson.edu.

[*] Corresponding author

**Abstract:** As quantum computing advances, the cryptographic algorithms that underpin confidentiality, integrity, and authentication in Intelligent Transportation Systems (ITS) face increasing vulnerability to quantum-enabled attacks. To address these risks, governments and industry stakeholders are turning toward post-quantum cryptography (PQC), a class of algorithms designed to resist adversaries equipped with quantum computing capabilities. However, existing studies provide limited insight into the implementation-focused aspects of PQC in the ITS domain. This review addresses that gap by evaluating the readiness of vehicular communication and security standards for adopting PQC. It examines in-vehicle networks and vehicle-to-everything (V2X) interfaces, and investigates vulnerabilities at the physical implementation layer of cryptographic hardware and embedded platforms, primarily exposure to side-channel and fault injection attacks. The review identifies thirteen research gaps: non-PQC-ready standards; constraints in embedded implementation and hybrid cryptography; interoperability and certificate-management barriers; a lack of real-world PQC deployment data in ITS; and physical-attack vulnerabilities in PQC-enabled vehicular communication. We present several future research directions, including updating vehicular communication and security standards, optimizing PQC for low-power devices, enhancing interoperability and certificate-management frameworks for PQC integration, conducting real-world evaluations of PQC-enabled communication and control functions across ITS deployments, and strengthening defenses against AI-assisted physical attacks. A phased roadmap is presented that aligns PQC deployment with regulatory, performance, and safety requirements, thereby guiding the secure evolution of ITS in the quantum computing era.

# 1. INTRODUCTION

## 1.1 Background and Motivation

The rapid evolution of connectivity and automation in transportation systems, characterized by the emergence of intelligent transportation systems (ITS), has significantly raised the importance of secure, resilient, and privacy-preserving data exchange. ITS refers to the integration of advanced information, communication, sensing, and intelligent decision-making technologies into transportation infrastructure and vehicles, aiming to enhance the efficiency, safety, and reliability of transportation networks [1,2]. It serves as a foundational framework in modern mobility by enabling seamless interaction between cyber and physical components across sectors, such as traffic control, fleet operations, transit security, and emergency response [3]. As these connected and autonomous vehicle (CAV) platforms rely increasingly on real-time communication and autonomous decision-making, cryptographic mechanisms have become essential to protect data integrity, confidentiality, and availability [4], commonly referred to as the CIA security triad as defined by the National Institute of Standards and Technology (NIST) [5]. Integrity ensures that transmitted ITS data and safety messages remain unaltered during transit, confidentiality protects sensitive vehicular and infrastructure information from unauthorized disclosure, and availability ensures that ITS communication and control services remain accessible and operational when needed for safe and reliable transportation operations. In addition, authenticity is a critical security requirement for ITS communications, ensuring that messages, certificates, and control commands originate from legitimate and trusted vehicles or infrastructure entities rather than malicious or spoofed sources.

According to the United States Department of Transportation's (USDOT) Architecture Reference for Cooperative and Intelligent Transportation (ARC-IT), cryptographic security is a fundamental requirement across ITS applications to ensure trusted data communication between vehicles, roadside units (RSUs) and infrastructure nodes [6]. While ARC-IT mainly focuses on cooperative vehicle-to-everything (V2X) communication, covering information exchanges via vehicle-to-vehicle (V2V), vehicle-to-infrastructure (V2I), vehicle-to-cloud (V2C), vehicle-to-pedestrian (V2P), vehicle-to-network (V2N), and infrastructure-to-infrastructure (I2I) communications, the broader ITS ecosystem also relies on secure in-vehicle and original equipment manufacturer (OEM) or cloud backhaul communications that connect with these V2X domains, helping maintain a continuous chain of trust and data protection across the entire transportation communication stack. As illustrated in **Fig. 1**, cryptographic mechanisms are essential for securing information exchanges across three primary ITS communication links, i.e., in-vehicle networks (e.g., Controller Area Network [CAN] or Automotive Ethernet secured for intra-vehicle message integrity), OEM/cloud backhaul channels (e.g., telematics and over-the-air [OTA] update channels protected through encrypted and authenticated sessions), and V2X/I2I interfaces (e.g., Basic Safety Message [BSM] or Signal Phase and Timing [SPaT] messages exchanged via IEEE 1609.2-based security frameworks [7]).

The communication links illustrated in **Fig. 1** represent several primary ITS domains in which cryptographic protections are applied to ensure secure communication across vehicles, infrastructure, and backend systems. In current deployments, these protections are predominantly built upon Elliptic Curve Cryptography (ECC)-based schemes standardized in IEEE 1609.2 and the European Telecommunications Standards Institute Cooperative ITS (ETSI C-ITS) security frameworks [7,8], providing efficient key management and digital signatures for vehicular communication. However, when such protections are weakened or compromised, adversaries could manipulate traffic signals, impersonate legitimate vehicles, or extract sensitive operational data, ultimately jeopardizing safety and public trust in ITS technologies.

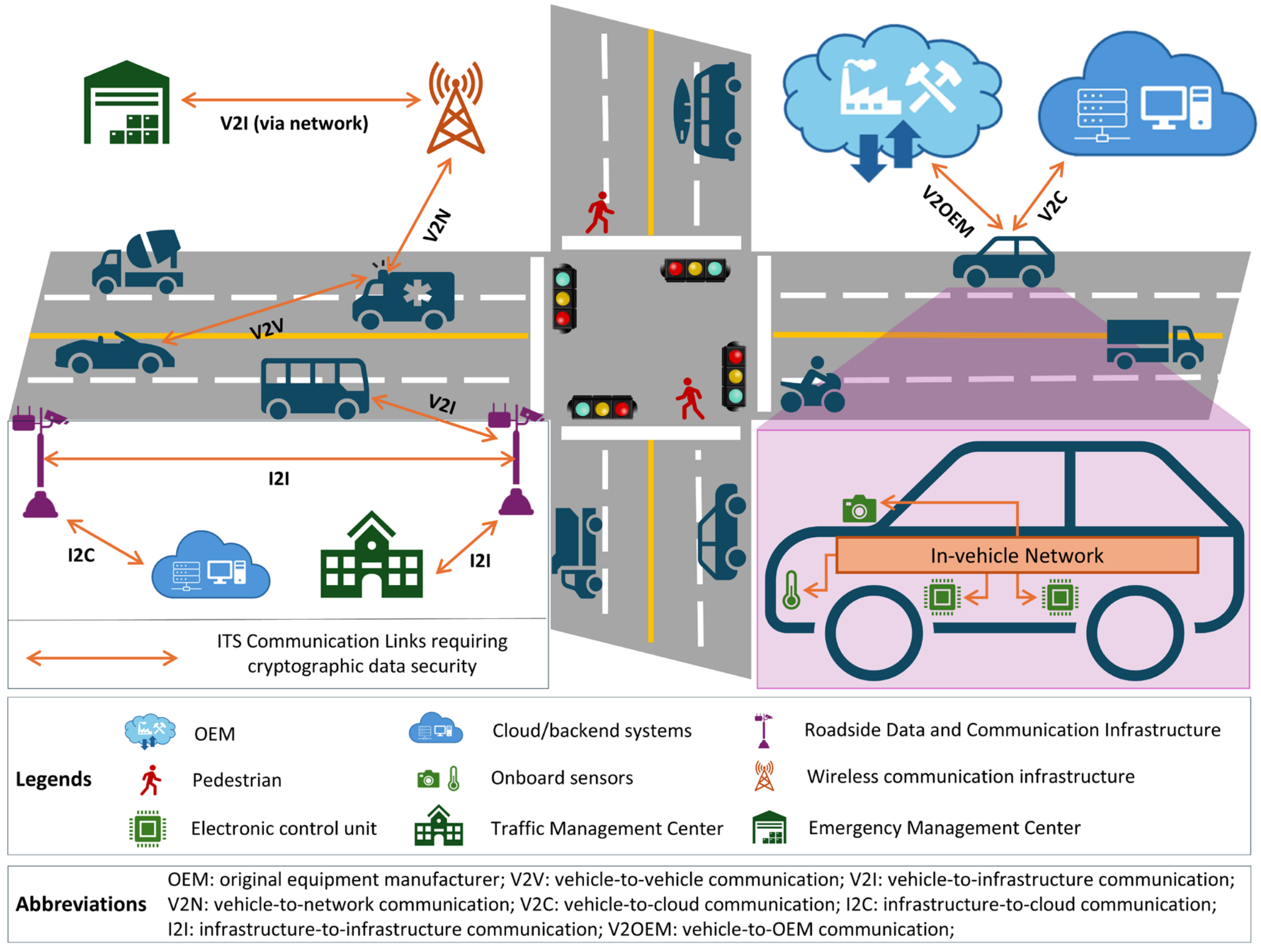

**Fig. 1.** Key ITS communication links that require cryptographic protection, including in-vehicle, OEM/cloud, and V2X/I2I data exchanges.

With the advent of quantum computing, the cryptographic algorithms that form the basis of modern communication security systems are becoming increasingly vulnerable. Quantum computers leverage quantum superposition and entanglement, enabling algorithmic speedups that threaten the cryptographic protections securing today's digital infrastructure. Algorithms such as Shor's algorithm [9] can efficiently factor large numbers and solve discrete logarithm problems, directly compromising widely used schemes, such as Rivest-Shamir-Adleman (RSA) [10], Diffie-Hellman (DH) [11], Digital Signature Algorithm (DSA) [12], and ECC (e.g., Elliptic Curve Digital Signature Algorithm [ECDSA] [13]). Furthermore, Grover's algorithm can drastically reduce the complexity of brute-force attacks on symmetric key algorithms such as the Advanced Encryption Standard (AES) [14]. This paradigm shift underscores the urgent need to transition from traditional cryptography to post-quantum cryptography (PQC), which refers to a class of cryptographic algorithms based on mathematically hard problems that are believed to be resistant to both quantum and classical computing attacks [15]. The urgency of this transition is particularly evident in the transportation sector, where vehicles and infrastructure typically operate for long periods, making them susceptible to "harvest-now-decrypt-later" attacks, in which encrypted data intercepted today could be decrypted once large-scale quantum computers become available [16,17].

In response to the emerging quantum threats, both governmental and industrial stakeholders have initiated coordinated efforts to accelerate the transition toward quantum-resistant cryptographic solutions (PQC). Recognizing the urgency of this transition, the U.S. federal government issued an Executive Order in January 2025 mandating that federal agencies strengthen cybersecurity by integrating quantum-resistant algorithms [18]. The Cybersecurity and Infrastructure Security Agency (CISA) has reinforced this mandate by urging the incorporation of PQC into government procurement processes, emphasizing its strategic importance for national infrastructure [19]. These efforts align with the ongoing standardization effort led by NIST, which culminated in the publication of the first set of PQC standards in 2024: Module-Lattice-Based Key Encapsulation Mechanism [ML-KEM (based on CRYSTALS-Kyber)] [20], Module-Lattice-Based Digital Signature Algorithm [ML-DSA (based on CRYSTALS-Dilithium)] [21], and Stateless Hash-based Digital Signature Algorithm [SLH-DSA (based on SPHINCS+)] [22], covering public-key encryption and digital signatures. These standards serve as the foundation for implementing PQC across critical sectors, including ITS. The private sector has also taken proactive steps, with leading technology companies, including Microsoft [23], Google [24], Apple [25], Amazon [26], Samsung [27], and Zoom [28], beginning to deploy PQC-enhanced encryption across operating systems, messaging services, cloud key infrastructure, and mobile platforms. Together, these initiatives indicate a unified effort to accelerate the deployment of quantum-resistant cryptographic standards across sectors, including the ITS sector.

While the NIST PQC standardization effort defines the cryptographic primitives (i.e., KEMs and digital signatures) needed for quantum-resilient security, their direct deployment into vehicular communication ecosystems raises unique system-level challenges that are not addressed by generic PQC evaluations. In this manuscript, "system-level" refers to considerations arising from integrating and operating PQC mechanisms within the end-to-end vehicular communication and trust ecosystem, including protocols, certificate and trust infrastructures, network timing and payload limits, embedded hardware platforms, and safety-critical operational workflows. Vehicular networks impose strict latency bounds, message-size limits, certificate lifecycle constraints, and safety-critical trust models that fundamentally shape how PQC algorithms are to be integrated into real-world ITS deployments.

### 1.2 Contributions and Focus of this Study

While there is growing recognition of the need for PQC in critical infrastructure, existing studies addressing PQC adoption in ITS remain limited in scope. **Table 1** summarizes a comparative analysis of recent survey efforts, demonstrating that this paper offers a broader and more integrated perspective by combining protocol-level analysis, implementation benchmarking, and deployment-centric discussion within the ITS ecosystem.

**Table 1**
**Comparative scope of recent surveys on PQC integration in ITS with respect to the present study**

| Study | Integrated In-vehicle-OEM-V2X View | Protocol-Level PQC Integration | Regulatory and Migration Alignment | Embedded Deployment Challenges | Structured ITS Research Gaps | Implementation Awareness and Benchmarking Insights | Roadmap and Deployment Strategy |
|---|---|---|---|---|---|---|---|
| Shim (2022) [29] | NP | NP | PP | P | P | NP | NP |
| Lonc et al. (2023) [30] | PP | P | P | PP | PP | NP | NP |
| Yoshizawa and Preneel (2023) [31] | NP | PP | P | NP | PP | NP | NP |
| Lohmiller et al. (2025) [16] | P | PP | PP | P | P | PP | NP |
| **Our Paper (2025)** | **P** | **P** | **P** | **P** | **P** | **P** | **P** |

Notes- P: Present; PP: Partially Present; NP: Not Present.

Several recent surveys have addressed specific aspects of PQC adoption within the transportation or vehicular communication domain. For example, Shim (2022) [29] focuses on PQC schemes, primarily detailing algorithmic structures and implementation trade-offs across different cryptographic families. In contrast, Lohmiller et al. (2025) [16] provide a broader and more structured survey of PQC migration across key vehicular domains, including in-vehicle networks, OEM-specific communications, and V2X protocols, and identify technical and regulatory challenges in each domain. However, their treatment is primarily descriptive, lacking systematic evaluation of integration-level trade-offs, benchmarking data, or in-depth analysis of implementation-layer constraints, such as latency, memory footprint, and platform-specific bottlenecks. Similarly, Lonc et al. (2023) [30] and Yoshizawa and Preneel (2023) [31] offer more focused discussions of certificate formats, message fragmentation, and migration barriers in V2X communications, yet neither study fully contextualizes these within the broader ITS architectural and regulatory landscape. Notably, none of these prior surveys provides a phased or implementation-focused roadmap to guide the real-world adoption of PQC across ITS communication links.

To address these shortcomings, this paper provides an implementation-focused survey that systematically investigates the integration of PQC across the entire ITS ecosystem. Accordingly, this paper focuses on the security and feasibility of PQC within vehicular communication protocols, spanning in-vehicle networks, OEM backhaul links, and V2X messaging and certificate infrastructures, rather than on the cryptographic design of PQC algorithms themselves. Unlike prior studies, this review adopts a holistic approach that contextualizes the vulnerabilities of traditional cryptographic schemes and the PQC standardization landscape, including the NIST initiative, to motivate the need for a post-quantum transition in safety-critical ITS communication.

The paper then builds on this contextual background to address practical deployment considerations, examining how PQC integrates with existing vehicular communication and security standards while evaluating computational, communication, and hardware implications under real-world ITS constraints. It also considers emerging implementation-level (i.e., considerations arising from the software and hardware realization of PQC algorithms) vulnerabilities of PQC, including susceptibility to side-channel, fault-injection, and profiling-based attacks, which are particularly relevant in resource-constrained automotive embedded systems. The paper concludes by synthesizing thirteen research gaps, outlining corresponding future research directions, and presenting a phased roadmap to guide the practical deployment of

PQC across ITS domains. In summary, the primary contributions of this paper are:

- An integrated analysis of how threats from quantum computers and emerging PQC standards impact the security foundations of vehicular communication systems, establishing the motivation for a post-quantum transition in ITS.
- An implementation-focused analysis that evaluates PQC integration across ITS communication links, addressing computational, communication, and hardware constraints in in-vehicle, OEM/cloud, and V2X/I2I environments.
- A forward-looking discussion supported by thirteen structured research gaps (RG1-RG13) that identify technical, regulatory, and deployment-level pathways toward real-world PQC implementation in ITS.

Each domain-specific section concludes with a synthesis that connects cryptographic properties to vehicular security implications, ensuring a system-level perspective across in-vehicle, OEM, and V2X communication domains. To avoid redundancy with existing NIST reports and PQC surveys, background discussion on quantum algorithms and PQC families is limited to what is necessary for interpreting vehicular deployment constraints.

## 2. REVIEW APPROACH

This section outlines the structured framework adopted to conduct a comprehensive review of PQC integration within ITS. The approach was designed to provide comprehensive coverage of PQC developments and to support a critical evaluation of PQC integration feasibility within ITS that led to the identification of emerging research challenges.

### 2.1 Literature Search and Inclusion Criteria

A total of 183 publications (2015-2026) were reviewed, encompassing peer-reviewed journal articles, conference papers, standards and regulatory documents, preprints and technical research reports, book chapters, and selected industry and consortium technical sources. Sources were identified through leading academic databases and publisher platforms (e.g., IEEE Xplore, ACM Digital Library, Elsevier, Springer, Wiley), major standards and government repositories (e.g., NIST, ETSI, ISO, UNECE), and cryptographic research archives (e.g., IACR ePrint, arXiv). To ensure comprehensive and representative coverage of PQC research relevant to ITS, the review prioritized sources that offered technical relevance, empirical validation, and contemporary insights, capturing both pre-standardization investigations and post-NIST developments. **Table 2** summarizes the literature categories and the distribution of sources included in this review.

**Table 2**
**Summary of literature categories**

| Category | Included Source Types | Count |
|---|---|---|
| Journal Articles | All journals (IEEE, ACM, Elsevier, Springer, Wiley, Nature, etc.) | 58 |
| Conference Papers | IEEE, ACM, Springer, LNCS, NDSS, CRYPTO, EUROCRYPT, VTC, ISCAS, DAC, HOST, ARES, etc. | 33 |
| Standards and Regulatory Documents | IEEE, ISO, ETSI, NIST FIPS/SP/IR, SAE, UNECE regulations, government cybersecurity standards | 38 |
| Preprints and Technical Research Reports | IACR ePrint, arXiv, theses, dissertations, non-archival technical manuscripts | 23 |
| Books and Book Chapters | Textbooks, edited volumes, encyclopedia chapters | 14 |
| Industry, Consortium and Web Technical Sources | Vendor white papers, corporate roadmaps, algorithm/project sites, standards portals, security blogs, technical documentation pages | 17 |

Notes- ACM: Association for Computing Machinery; ARES: International Conference on Availability, Reliability and Security; AWS: Amazon Web Services; CRYPTO: International Cryptology Conference; DAC: Design Automation Conference; ETSI: European Telecommunications Standards Institute; EUROCRYPT: International Conference on the Theory and Applications of Cryptographic Techniques; FIPS: Federal Information Processing Standards; HOST: IEEE International Symposium on Hardware Oriented Security and Trust; IACR: International Association for Cryptologic Research; IEEE: Institute of Electrical and Electronics Engineers; IR: Interagency Report; ISCAS: IEEE International Symposium on Circuits and Systems; ISO: International Organization for Standardization; LNCS: Lecture Notes in Computer Science (Springer); NDSS: Network and Distributed System Security Symposium; NIST: National Institute of Standards and Technology; NSA: National Security Agency; SAE: Society of Automotive Engineers; SP: Special Publication; TCHES: IACR Transactions on Cryptographic Hardware and Embedded Systems; UN R155: United Nations Regulation No. 155 (Cybersecurity and Cybersecurity Management System); UN R156: United Nations Regulation No. 156 (Software Update and Software Update Management System); VNC: IEEE Vehicular Networking Conference; VTC: IEEE Vehicular Technology Conference.

## 2.2 Analytical Domains

The reviewed literature was organized into six analytical domains (D1-D6) representing key dimensions of PQC integration in vehicular environments:

- D1: Quantum-era cryptographic risk and PQC foundations;
- D2: Vehicular and communication standards guiding security architectures;
- D3: In-vehicle network integration and protocol-level constraints;
- D4: OEM-specific and cloud backhaul communications;
- D5: V2X communication security and hybrid migration strategies; and
- D6: Physical and implementation-level cryptographic attack threats.

These domains collectively represent the full progression of PQC deployment, progressing from quantum-era cryptographic risk and PQC foundations (D1) through standardization and regulatory alignment (D2) to practical integration (D3-D5) and implementation resilience (D6). **Fig. 2** summarizes this progression by illustrating how the review framework transitions through these domains, beginning with the quantum-vulnerability context and PQC standardization background, followed by the evaluation of current vehicular communication and security standards, the assessment of PQC integration feasibility across ITS communication links, the investigation of physical-layer attacks on PQC in automotive hardware, and the synthesis of emerging research gaps and future directions.

**Table 3** presents the scope and distribution of the literature across each domain, including overlapping counts where studies relate to multiple domains. Notably, none of the 18 studies classified under D2 (Vehicular Standards) are exclusive to that domain; each intersects with at least one area. This underscores the central role that standards play in the broader PQC landscape, bridging foundational PQC research and standardization efforts (D1) with implementation and system-level studies (D3-D6).

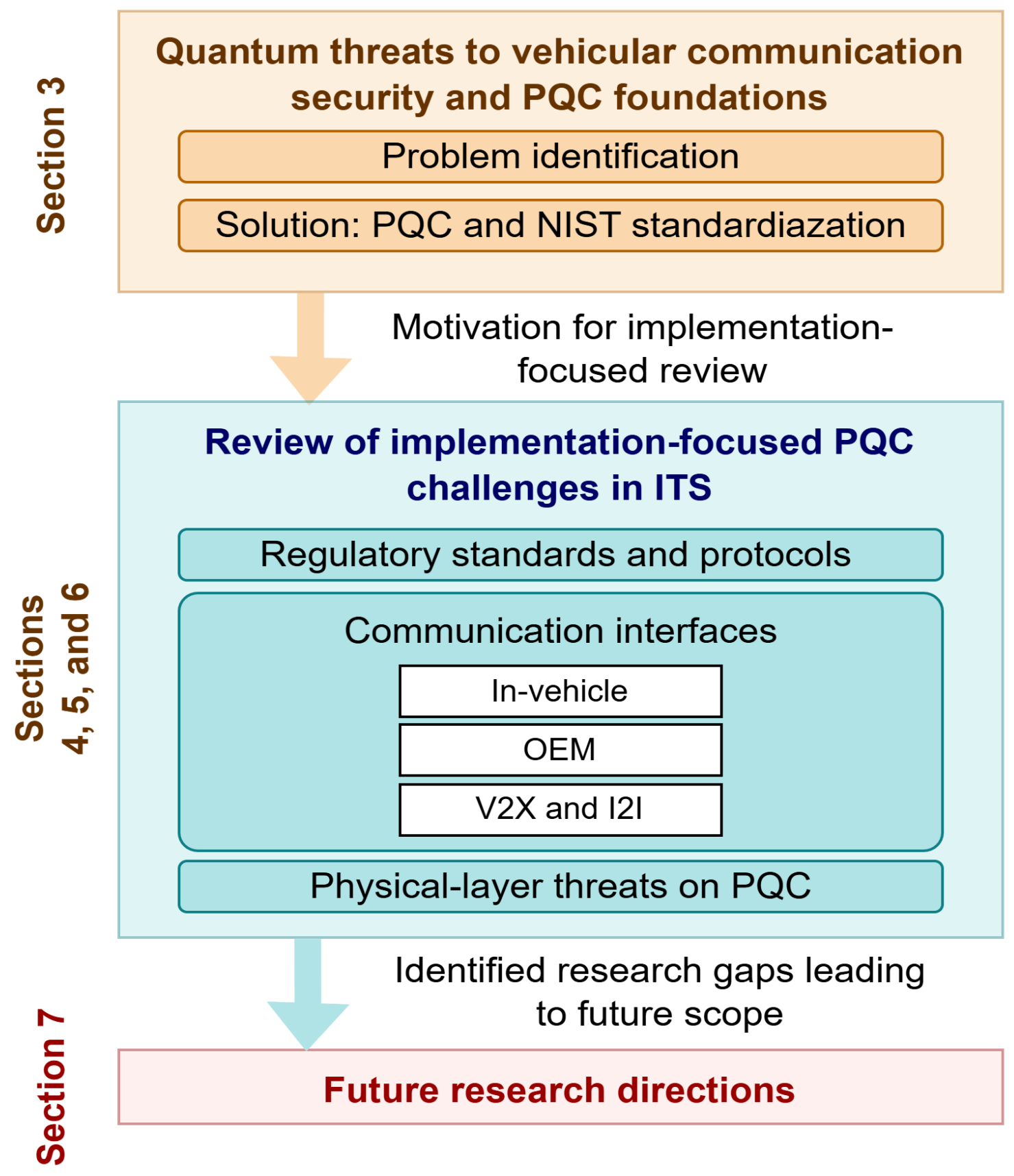

**Fig. 2.** Overview of the review progression, beginning with classical cryptographic vulnerabilities and NIST PQC standardization; followed by identifying research gaps related to PQC adoption in current vehicular communication and security standards, PQC integration feasibility across ITS communication links, and the susceptibility of PQC implementations to physical-layer attacks in automotive hardware; and concluding with corresponding future research directions. Abbreviations – OEM: Original Equipment Manufacturer; V2X: Vehicle-to-Everything; I2I: Infrastructure-to-Infrastructure.

**Table 3**
**Domain classification**

| Domain | Scope of the Survey | Count |
|---|---|---|
| D1 | Quantum-era cryptographic risk and PQC foundations, including quantum attack models, vulnerabilities of classical cryptography, PQC algorithm families, NIST standardization, and deployment-relevant security parameters of PQC | 59 |
| D2 | Vehicular communication, cybersecurity, and regulatory standards, with PQC-readiness and certificate/trust architecture analysis | 18 |
| D3 | Secure cryptographic integration in in-vehicle networks, including ECU platforms, automotive in-vehicle communication protocols, payload/timing limits, and embedded resource constraints | 30 |
| D4 | OEM backend, cloud backhaul, and OTA communication security, including lifecycle trust management, update pipelines, and backend cryptographic infrastructure | 12 |
| D5 | V2X and I2I communication security, including PKI/SCMS trust models, certificate handling, message authentication, and hybrid classical-PQC migration strategies | 50 |
| D6 | Physical and implementation-layer threats to PQC, including side-channel, fault-injection, and AI-assisted profiling attacks and corresponding countermeasures | 23 |

Notes- AI: Artificial Intelligence; ECU: Electronic Control Unit; I2I: Infrastructure-to-Infrastructure; NIST: National Institute of Standards and Technology; OEM: Original Equipment Manufacturer; OTA: Over-The-Air; PKI: Public Key Infrastructure; PQC: Post-Quantum Cryptography; SCMS: Security Credential Management System; V2X: Vehicle-to-Everything.

## 2.3 Research Gap Derivation Process

To ensure robustness and reliability in identifying research gaps, a conservative cross-validation approach was employed. Specifically, a research gap was formally documented only if it was independently confirmed by at least three distinct sources across the reviewed literature. This threshold was chosen to reduce bias from isolated findings and to ensure that each documented research gap reflects a recurring and validated concern within the PQC and ITS research community. Using this approach, thirteen research gaps (RG1-RG13) were identified and are presented later in detail. Gaps associated with standardization (D2) were the most prominent, particularly regarding certificate structures, key management, and interoperability readiness. Implementation-focused domains (D3-D6) revealed additional challenges related to computational and communication latency, memory usage, and susceptibility to side-channel or fault-injection attacks in embedded systems. Together, these insights informed the development of the research roadmap presented in **Section 7**, which identifies key technical and regulatory directions for advancing PQC deployment in transportation systems.

## 3. QUANTUM VULNERABILITIES OF CRYPTOGRAPHY USED IN VEHICULAR SYSTEMS AND PQC FOUNDATIONS FOR MIGRATION

This section explains why current cryptographic mechanisms used in vehicular communication, particularly ECC-based authentication in IEEE 1609.2 and ETSI C-ITS, become vulnerable under quantum computing, with emphasis on the practical impacts of Shor's and Grover's algorithms. It then translates these quantum risks into migration-relevant requirements for ITS, highlighting why preserving trust and message authenticity is the primary priority and why symmetric primitives can often be mitigated through conservative parameter selection. Finally, to support migration planning from quantum-vulnerable to quantum-safe primitives, it introduces the post-quantum design space (PQC primitives and standardization context) at a high level and only to the extent needed to evaluate deployment feasibility under ITS constraints, such as message size, latency, and certificate/fragmentation overhead.

### 3.1 Cryptographic Building Blocks for Digital Communication Security

Cryptographic schemes can be classified into several types based on their functions: public key or asymmetric cryptography, symmetric cryptography, digital signatures, and hashing functions [32]. Public key cryptography uses a pair of keys: a public key, which can be shared openly, and a private key, which is kept secret. It is primarily used for secure key exchange and digital signatures, with examples including RSA, DH, and ECC [32]. Symmetric cryptography, such as the AES, uses a single key for both encryption and decryption processes, making it particularly efficient for encrypting large datasets [32]. However, this secret key must be securely communicated between parties before encryption and decryption can occur, often requiring additional mechanisms or protocols for key exchange, such as key encapsulation mechanisms (KEMs) [33]. Digital signatures, such as ECDSA and DSA, verify the authenticity and integrity of a message, providing non-repudiation [32]. Hashing functions generate a fixed-size hash value from input data, ensuring data integrity verification and certificate management in secure communication systems such as SHA-256 [32]. Together, these primitives form the foundation of modern digital and internet security.

Although traditional cryptographic primitives have provided robust protection for decades, their security rests on the assumption that hard mathematical problems, such as integer factorization and discrete logarithms, are infeasible for classical computers to solve. The emergence of quantum computers fundamentally challenges this assumption, as large-scale, fault-tolerant quantum computers equipped with quantum algorithms could solve these problems exponentially faster, undermining the confidentiality, authenticity, and integrity of digital communications. Before delving into quantum threats and post-quantum migration in an ITS context, we present key quantum-computing concepts referenced throughout this paper in **Table 4**.

**Table 4**
**Key quantum-computing concepts relevant to cryptographic security**

| Concept | Description |
|---|---|
| Qubit | Fundamental unit of quantum information. Unlike a classical bit (0 or 1), a qubit can exist in a superposition of 0 and 1, enabling exploration of exponentially many states. |
| Superposition | Quantum state in which a qubit simultaneously represents multiple values between 0 and 1, allowing quantum computers to evaluate many possibilities at once. |
| Entanglement | A correlated quantum state between two or more qubits such that the state of one qubit instantaneously influences the others, enabling parallel operations. |
| Physical Qubit | A qubit implemented in actual hardware (e.g., superconducting circuits, trapped ions). Physical qubits suffer from noise, decoherence, and operational errors. |
| Logical Qubit | An error-corrected abstraction built from multiple physical qubits. Logical qubits enable reliable computation but require substantial hardware overhead. |
| Quantum Error Correction (QEC) | Techniques that detect and correct errors in qubits' states, which are necessary for maintaining long-running computations but significantly increasing qubit requirements. |
| Quantum Gates | Operations that manipulate qubit states (analogous to classical logic gates). Examples include: Hadamard (creates superposition) and CNOT (generates entanglement). |
| Gate Depth | Number of sequential gate operations in a quantum circuit. Higher depth increases runtime and error accumulation. |
| MAXDEPTH | Maximum feasible depth of a quantum circuit (i.e., the number of sequential quantum gate layers) that can be executed reliably on a fault-tolerant quantum computer before accumulated errors dominate. |
| Fault-Tolerant Quantum Computer | A system capable of running long, complex quantum algorithms using logical qubits and active QEC. Represents the milestone at which quantum attacks on classical cryptography become practically feasible. |

## 3.2 Quantum Threats and Why It Matters for ITS

### *3.2.1 Cryptographic Impact of Shor's and Grover's Algorithms*

Two quantum algorithms play a central role in the emerging threat to classical cryptographic systems. Shor's algorithm [9] efficiently factors large integers and computes discrete logarithms, directly compromising widely used public-key schemes, such as RSA, DH, DSA, and ECC [34–38]. Early analyses by Proos and Zalka [39] estimated that factoring a 1024-bit RSA key would require roughly 2,000 logical qubits, while breaking 160-bit ECC would need about 1,000 logical qubits. More recent resource estimates refined these results, indicating that factoring a 2048-bit RSA key could require approximately 20 million physical qubits and eight hours of runtime on a fault-tolerant quantum computer [40] or approximately 1-2 million physical qubits under optimized circuit depth [36], whereas 256-bit ECDSA could be broken in about one day using around 13 million physical qubits [41].

Grover's algorithm [14] targets symmetric-key cryptography and hashing functions by offering a quadratic speedup in brute-force search operations, effectively halving the key strength of classical systems. Under this model, AES-128, which uses a 128-bit key, offers only about 64-bit post-quantum security, while AES-256 maintains an effective strength comparable to classical AES-128 [15]. SM4, another 128-bit symmetric cipher widely adopted in vehicular standards, is subject to the same key-search speedup from Grover's algorithm, reducing its effective post-quantum security level to approximately 64 bits [42]. However, practical feasibility remains constrained by circuit-depth limits, error correction, and limited parallelization. Resource estimates by Grassl et al. [43] and analyses by the UK National Cyber Security Centre (NCSC) [44] indicate that realistic implementations of Grover-style brute force remain extremely expensive; for example, scaled cycle costs remain very high (approximately $2^{128.7}$ for AES-128 and $2^{257.7}$ for AES-256 under realistic assumptions) [45]. These overheads suggest that quantum brute-force attacks on AES remain computationally expensive in practice, although the long-term security margin of smaller key sizes is reduced under the quadratic speedup model.

Similar considerations apply to hash functions, where Grover's algorithm can significantly accelerate the process of finding collisions in which two different inputs result in the same hash output, thereby reducing their security. Specifically, using Grover's algorithm, a hash function with an n-bit output requires approximately $2^{n/2}$ operations to break preimage resistance, which is the property that makes it computationally infeasible to determine the original input from a given hash output [14]. Furthermore, a related technique known as the quantum birthday attack, which combines Grover's algorithm with the classical birthday paradox, can find collisions even more efficiently [45]. To counter these threats, hash functions require substantially longer outputs in the quantum era. For instance, according to Brassard et al., a hash function aiming for b-bits of security against Grover's attacks should produce a 3b-bit output [46]. While many current hash algorithms fall short of this requirement, SHA-2 and SHA-3, with longer output sizes (at least 256-bit output), offer greater resistance to quantum attacks [15]. According to NIST's transition guideline, SHA3-512, due to its 512-bit output and 256-bit collision resistance, remains a robust option for post-quantum integrity protection and message authentication [47].

### *3.2.2 System-Level Consequences for Vehicular Trust and Safety*

The vulnerabilities from Shor's and Grover's algorithms have direct implications for current vehicular communication security standards, which continue to rely on quantum-vulnerable cryptographic primitives. The IEEE 1609.2 standard for Wireless Access in Vehicular Environments (WAVE) specifies algorithms to ensure the integrity, confidentiality, and authenticity of messages between vehicles, RSUs, and backend servers [7]. These include ECDSA for digital signatures [13], AES-128 [48] and SM4 [49] for symmetric encryption, Elliptic Curve Integrated Encryption Scheme (ECIES) [50] and SM2 [51] for key establishment, and hash functions such as SHA-256 [52] and SM3 [53] for data integrity. Similarly, ETSI C-ITS security standards adopt ECDSA for digital signatures and SHA-256 for hashing to secure ITS communications [8,54]. Shor's algorithm compromises asymmetric schemes, such as ECDSA, ECIES, and SM4, while Grover's algorithm reduces the effective strength of symmetric and hash-based primitives like AES-128, SM4, and SHA-256. Together, these primitives form the vehicular trust infrastructure that enables authenticated safety messaging, certificate validation, and secure OTA and backhaul communications.

From a quantum-attack perspective, Shor's algorithm directly compromises the asymmetric and signature layer that anchors vehicular trust, including ECC-based digital signatures and key-establishment mechanisms, such as ECDSA, ECIES, and SM2, making impersonation, forged safety messages, and malicious software updates feasible once large-scale quantum computers exist. In contrast, Grover's algorithm reduces the effective security of symmetric encryption and hash functions, including AES-128, SM4, SHA-256, and SM3, by roughly a square root, but this degradation can generally be mitigated by increasing key and output sizes (e.g., AES-256 and SHA-512) without redesigning protocols. Consequently, the primary cryptographic risk for ITS is the breakdown of ECC-based authentication and key establishment mechanisms under Shor's algorithm, while Grover's algorithm mainly creates a parameter-strengthening requirement, not a structural break in the ITS security architecture, including certificates, digital signatures, and key management used in vehicular communication systems.

The impact of quantum computing on various traditional cryptographic schemes is further summarized in **Table 5**. This table is adapted from NIST and illustrates how quantum computing compromises the security of traditional cryptographic algorithms, making schemes such as ECC, RSA, and DSA no longer secure [15]. Given the vulnerabilities, it is evident that current cryptographic schemes, particularly those incorporated in vehicular standards, are insufficient against the advancing capabilities of quantum computers. The next subsection examines the expected progress of quantum hardware to contextualize when these cryptanalytic risks may become practically relevant for deployed vehicular systems.

**Table 5**
**Impact of quantum computing on traditional cryptographic schemes**

| Cryptographic Scheme | Type | Purpose | Quantum Vulnerability Mechanism | Implication of Large-Scale Quantum Computer | References |
|---|---|---|---|---|---|
| AES | Symmetric Key | Encryption | Quadratic speedup (Grover's algorithm) | Larger key sizes needed (e.g., replacing AES-128 with AES-256) | [14,15] |
| SHA-2, SHA-3 | Hash Functions | Hashing | Collision search speedup (Grover's algorithm) | Longer output needed (e.g., replacing SHA-256 with SHA-384 or SHA-512) | [14,15] |
| RSA | Public Key | Signatures, Key Establishment | Integer factorization (Shor's algorithm) | No longer secure | [9,15] |
| ECDSA, ECDH (ECC) | Public Key | Signatures, Key Exchange | Elliptic curve discrete logarithm (Shor's algorithm) | No longer secure | [9,15] |
| DSA (Finite Field Cryptography) | Public Key | Signatures, Key Exchange | Discrete logarithm (Shor's algorithm) | No longer secure | [9,15] |

Notes- AES: Advanced Encryption Standard; DSA: Digital Signature Algorithm; ECC: Elliptic Curve Cryptography; ECDH: Elliptic Curve Diffie-Hellman; ECDSA: Elliptic Curve Digital Signature Algorithm; RSA: Rivest-Shamir-Adleman; SHA: Secure Hash Algorithm.

## 3.3 Quantum-Hardware Progress and Cryptographic Migration Timeline

The transition to PQC in ITS is driven not by speculative projections, but by the recognition that quantum-capable cryptanalytic attacks may become feasible within the multi-decade operational lifetime of vehicular systems. Government and standards bodies have emphasized that cryptographic systems deployed today must remain secure against future quantum adversaries [15,18,19,47], particularly given the feasibility of 'harvest-now-decrypt-later' and 'forge-later' attacks against long-lived infrastructure and archived vehicular data. Because vehicles, RSUs, and transportation backend systems are often deployed for 15-30 years, cryptographic decisions made today must account for exposure to future fault-tolerant quantum computers.

Against this backdrop, commercial quantum-hardware roadmaps provide additional contextual evidence that steady progress toward large-scale, error-corrected quantum computing is ongoing. Major technology developers have publicly outlined paths toward machines with hundreds to thousands of logical qubits and large-depth quantum circuits. For example, IBM's roadmap targets systems capable of very large circuit depths and on the order of a few hundred logical qubits around 2029, with further multibillion-gate scale expansion projected in the early 2030s [55]. Microsoft's staged transition from physical qubits to logical qubits, and then to scalable quantum supercomputers, includes milestone targets such as approximately one million reliable quantum operations per second in the late 2020s to early 2030s timeframe [56,57]. Google has outlined plans for large-scale error-corrected systems with up to about one million physical qubits within a similar late-2020s/early-2030s horizon [58]. IonQ has projected that modular trapped-ion architectures will scale to more than two million physical qubits (tens of thousands of logical qubits) by around 2030 [59]. While these roadmaps are not guarantees of cryptanalytic capability on any specific timeline, they collectively illustrate that sustained investment and engineering progress toward quantum-capable platforms is underway across multiple hardware paradigms.

For ITS security, the exact year in which Shor-scale attacks become practical is less important than the fact that they may occur within the lifetime of deployed vehicular infrastructure. This uncertainty, combined with the safety-critical role of cryptographic authentication in V2X communication, makes early migration to post-quantum-safe digital signatures and key-establishment mechanisms an engineering requirement. Accordingly, post-quantum readiness needs to be incorporated into vehicular standards, certificate infrastructures, and communication protocols well before quantum computers reach full cryptanalytic maturity. The following subsections, therefore, introduce PQC as the primary migration pathway and summarize the standardized PQC primitives and security levels relevant to vehicular systems.

### 3.4 Post-Quantum Cryptography for Secure Vehicular Communications

Post-quantum cryptography (PQC) refers to cryptographic algorithms developed to withstand attacks from both classical and quantum computers by using mathematical problems that are believed to remain computationally hard in the quantum era. [15]. Unlike quantum-key distribution or hardware-dependent approaches, PQC algorithms are software-compatible with existing communication stacks, making them practical for incremental deployment across vehicles, RSUs, and backend systems.

NIST's standardization process provides the cryptographic building blocks for this transition. The first standardized PQC primitives published in August 2024 include ML-KEM (formerly CRYSTALS-Kyber) for key encapsulation [20], ML-DSA (formerly CRYSTALS-Dilithium) for digital signatures [21], and SLH-DSA (formerly SPHINCS+) for hash-based signatures [22]. Falcon remains in draft standardization [47], and HQC has been selected as an additional standardized KEM with a draft specification under preparation [60]. NIST has also issued guidance on the secure use of KEMs in both standalone and hybrid cryptographic systems [61] and continues a parallel effort to expand the set of post-quantum digital signature schemes [62,63].

To enable consistent comparison across candidate algorithms, NIST defines five security strength categories (Levels 1-5) based on the estimated computational effort required to compromise widely deployed symmetric primitives [64]. Specifically, Levels 1, 3, and 5 are defined by equivalence to the effort required to break AES-128, AES-192, and AES-256, respectively, while Levels 2 and 4 correspond to the effort required to break SHA3-256 and SHA3-384 under their standard security models [64]. These levels, summarized in **Table 6**, provide a standardized way to map PQC parameters to classical and quantum attack cost models. From the ITS perspective, these categories are critical because higher security levels translate directly into larger keys, longer signatures, and higher computational cost, which, in turn, determine the feasibility of real-time V2X messaging and certificate handling.

**Table 6**
**NIST security strength categories and computational requirements** [64]

| Security Strength Category | Reference security target (cost equivalence) | Estimated computational cost (quantum/classical) |
|---|---|---|
| Level 1 | Equivalent to breaking AES-128 | $2^{170}$/MAXDEPTH quantum gates or $2^{143}$ classical gates |
| Level 2 | Equivalent to breaking SHA3-256 | $2^{146}$ classical gates |
| Level 3 | Equivalent to breaking AES-192 | $2^{233}$/MAXDEPTH quantum gates or $2^{207}$ classical gates |
| Level 4 | Equivalent to breaking SHA3-384 | $2^{210}$ classical gates |
| Level 5 | Equivalent to breaking AES-256 | $2^{298}$/MAXDEPTH quantum gates or $2^{272}$ classical gates |

While NIST standardizes individual algorithms, vehicular deployment depends on which PQC families are practical under bandwidth, latency, and embedded-system constraints. **Table 7** compares the five major PQC families from this deployment-centric viewpoint. Lattice-based schemes (e.g., Kyber, Dilithium, Falcon) offer the best balance of efficiency, security, and artifact size [15,65]. Code-based schemes (e.g., HQC) provide strong security but often require very large public keys [66,67], hash-based schemes (e.g., SPHINCS+) rely on conservative assumptions but produce large signatures [66,68], and multivariate and isogeny-based schemes have suffered cryptanalytic breaks or maturity concerns [60,66,69–71]. In addition, the lattice-based family uniquely supports advanced primitives such as Fully Homomorphic Encryption (FHE) [72], enabling computation over encrypted vehicular data for privacy-preserving analytics and cloud-assisted ITS services [73,74]. These combined properties explain why lattice-based PQC is widely regarded as the most promising cryptographic foundation for vehicular communication standards.

**Table 7**
**Key strengths and weaknesses of PQC algorithmic families**

| PQC Family | Example Schemes | Strengths | Weaknesses |
|---|---|---|---|
| Lattice-Based | Kyber [20], Dilithium [21], and Falcon [75] | Efficient and flexible constructions with moderate key and signature sizes suitable for embedded and V2X systems | Some variants (e.g., Falcon) are complex to implement |
| Code-Based | Classic McEliece [67], and HQC [76] | Strong resistance to quantum attacks | Very large public key sizes, challenging for bandwidth-limited V2X and certificate distribution |
| Multivariate Polynomial | GeMSS [77] and Rainbow [78] | Fast signature generation | Cryptographically broken due to algebraic attacks; unsuitable for deployment |
| Hash-Based | SPHINCS+ [22] | Minimal reliance on unproven assumptions | Large signature sizes, increasing packet fragmentation and broadcast overhead |
| Isogeny-Based | SIKE [79] | Extremely small key and ciphertext sizes | Cryptographically broken by key-recovery attacks; no longer viable |

## 3.5 Why PQC Changes the Deployment Reality for ITS

Although PQC restores cryptographic security against Shor-based attacks, it introduces substantial system-level overheads that directly impact vehicular networks. These effects are evidenced by the cryptographic artifact sizes reported in **Table 8**, which lists the public-key sizes, ciphertext lengths, and signature sizes of standardized and candidate PQC schemes. Compared with ECC, post-quantum public keys, ciphertexts, and digital signatures are typically larger, often by one to two orders of magnitude, thereby increasing bandwidth consumption, packet fragmentation, and certificate distribution cost in wireless V2X channels.

This size inflation is most acute for the authentication layer, which is the core of vehicular trust. For example, according to IEEE 1609.2, ECC-based ECDSA (P-256) signatures are typically 64 bytes [7], whereas ML-DSA signatures are on the order of a few kilobytes (e.g., ~2.4-4.6 KB across parameter sets in **Table 8**), and SLH-DSA signatures can reach tens of kilobytes depending on the selected variant (as reflected by the SLH-DSA parameter sets in **Table 8**). Likewise, PQC KEMs introduce substantially larger artifacts (i.e., public/private keys, ciphertexts, and transmitted cryptographic payloads) than ECC-based key establishment. For example, ML-KEM ciphertexts and public keys are ~800 bytes to ~1.5 KB, while HQC artifacts are multiple kilobytes (**Table 8**), which can increase handshake payload size and retransmission risk on lossy wireless links. In IEEE 1609.2 and ETSI C-ITS, certificate chains, pseudonym rotation, and broadcast authentication require operating within strict packet-size and latency constraints [7,8,54]. As a result, larger PQC artifacts translate into practical integration challenges (e.g., fragmentation behavior, verification latency under load, and certificate-caching pressure) rather than a simple cryptographic substitution.

Beyond communication overhead, PQC can increase memory footprint and computational load on on-board units (OBUs) and RSUs that operate under real-time and power constraints. As the security-strength categories in **Table 6** increase, the corresponding parameter growth in **Table 8** can increase verification time and storage requirements, affecting safety-message verification latency, certificate distribution and caching strategy, and secure OTA and OEM backhaul workflows.

**Table 8**
**Claimed security levels and key sizes of PQC schemes standardized or currently under consideration for standardization (adapted from [20–22,60,66])**

| PQC Candidate | Claimed NIST Security Category | Public Key (in bytes) | Private Key (in bytes) | Ciphertext/ Signature (in bytes) | Shared Secret Key (in bytes) | Core SVP Estimate | Gate Count | Memory |
|---|---|---|---|---|---|---|---|---|
| **[a]ML-KEM -512** | **Level 1** | **800** | **1632** | **768** | **32** | **C:118 bits; Q:107 bits** | $2^{151}$ | $2^{94}$ |
| **[a]ML-KEM-768** | **Level 3** | **1184** | **2400** | **1088** | **32** | **C:183 bits; Q:166 bits** | $2^{215}$ | $2^{139}$ |
| **[a]ML-KEM-1024** | **Level 5** | **1568** | **3168** | **1568** | **32** | **C:256 bits; Q:232 bits** | $2^{287}$ | $2^{190}$ |
| [b]HQC-128 | Level 1 | 2249 | 40 | 4481 | 32 | NA | NA | NA |
| [b]HQC-192 | Level 3 | 4522 | 40 | 9026 | 32 | NA | NA | NA |
| [b]HQC-256 | Level 5 | 7245 | 40 | 14469 | 32 | NA | NA | NA |
| **[a]ML-DSA-44** | **Level 2** | **1312** | **2560** | **2420** | **NA** | **C:123 bits; Q:112 bits** | $2^{159}$ | $2^{98}$ |
| [a]ML-DSA-65 | Level 3 | 1952 | 4032 | 3309 | NA | C:182 bits; Q:165 bits | $2^{217}$ | $2^{139}$ |
| [a]ML-DSA-87 | Level 5 | 2592 | 4896 | 4627 | NA | C:252 bits; Q:229 bits | $2^{285}$ | $2^{187}$ |
| **[b]Falcon-512** | **Level 1** | **897** | **7553** | **666** | **NA** | **C:120 bits; Q:108 bits** | **NA** | **NA** |
| **[b]Falcon-1024** | **Level 5** | **1793** | **13953** | **1280** | **NA** | **C:273 bits; Q:248 bits** | **NA** | **NA** |
| [a]SLH-DSA-SHA2/SHAKE -128s | Level 1 | 32 | 64 | 7856 | NA | NA | NA | NA |
| [a]SLH-DSA-SHA2/SHAKE -128f | Level 1 | 32 | 64 | 17088 | NA | NA | NA | NA |

| PQC Candidate | Claimed NIST Security Category | Public Key (in bytes) | Private Key (in bytes) | Ciphertext/ Signature (in bytes) | Shared Secret Key (in bytes) | Core SVP Estimate | Gate Count | Memory |
|---|---|---|---|---|---|---|---|---|
| [a]SLH-DSA-SHA2/SHAKE -192s | Level 3 | 48 | 96 | 16224 | NA | NA | NA | NA |
| [a]SLH-DSA-SHA2/SHAKE -192f | Level 3 | 48 | 96 | 35664 | NA | NA | NA | NA |
| [a]SLH-DSA-SHA2/SHAKE -256s | Level 5 | 64 | 128 | 29792 | NA | NA | NA | NA |
| [a]SLH-DSA-SHA2/SHAKE -256f | Level 5 | 64 | 128 | 49856 | NA | NA | NA | NA |

Notes- [a]Standardized by NIST on August 13, 2024; [b]Standardization process ongoing; C: Classical; Q: Quantum; SVP: Short Vector Problem; NA: Not Available; Bold rows indicate NIST-standardized PQC schemes and primary deployment candidates for near-term ITS integration, considering both security strength and ITS performance constraints.

## 3.6 Concluding Remarks of Section 3

This section establishes the cryptographic motivation for PQC migration in ITS by connecting quantum algorithm capabilities, emerging hardware roadmaps, and the dependence of vehicular trust infrastructures on quantum-vulnerable asymmetric primitives. The synthesis shows that the primary system-level risk for ITS arises from the potential collapse of ECC-based authentication and key-establishment layers, whereas symmetric primitives can often be strengthened by scaling their parameters. It also explained why PQC selection in transportation contexts should be evaluated not only by theoretical security but also by artifact size, verification cost, and deployment compatibility. However, this review did not perform an independent cryptanalytic reassessment of PQC hardness assumptions or quantum resource estimates and instead relied on published standards reports and peer-reviewed analyses. In addition, detailed mathematical constructions of PQC schemes were intentionally out of scope, since the focus here is on deployment implications rather than algorithm design. The next section examines how these cryptographic requirements translate into current vehicular communication and cybersecurity standards and evaluates their readiness for PQC integration.

# 4. CURRENT STANDARDS FOR VEHICULAR COMMUNICATION SECURITY

The security of modern vehicular and infrastructure communications, including in-vehicle networks, OEM-specific services, and V2X interactions, is supported by a diverse set of cryptographic standards developed by regulatory, industry, and regional bodies. These standards ensure message integrity, confidentiality, and authenticity across vehicles, RSUs, and backend servers through classical cryptographic primitives, such as ECDSA and ECIES.

However, these standards were designed before the emergence of practical quantum threats and therefore do not support PQC algorithms. As a result, current standards need to be examined for their compatibility with future quantum-resistant security frameworks. This section describes the primary security standards used in vehicular and infrastructure communication, their reliance on ECC and public key infrastructure (PKI)-based trust models, and evaluates their current level of PQC readiness.

### 4.1 IEEE 1609.2 (WAVE Security)

IEEE 1609.2 is a U.S. standard that defines secure message formats and a vehicular PKI (VPKI) for V2X communications, particularly over Dedicated Short-Range Communications (DSRC) [7,80,81]. It mandates the use of ECC, specifically ECDSA digital signatures using the NIST P-256 curve, to authenticate safety messages [80]. Certificates under IEEE 1609.2 can be either explicit, where the full public key is included, or implicit, where the public key is derived using the Elliptic Curve Qu-Vanstone technique with assistance from a certificate authority (CA), thereby reducing the certificate size. The underlying trust model uses a hierarchical structure, with a Root CA delegating to subordinate authorities that issue short-lived pseudonym certificates to vehicles. These ephemeral certificates enhance privacy by frequently rotating identifiers. Data encryption, where applicable, utilizes ECIES in conjunction with symmetric AES-CCM encryption [81]. Overall, the standard is optimized for DSRC/WAVE's bandwidth and latency constraints, and its security is tightly coupled with a compact ECC-based certificate infrastructure. Because IEEE 1609.2 certificate and message security formats are tightly optimized around compact ECC artifacts, direct substitution with PQC signatures and keys would exceed current size and fragmentation assumptions. Therefore, PQC integration into IEEE 1609.2 will require hybrid certificate models, format extensions, and fragmentation-aware processing, as further discussed in Section 5.

### 4.2 ETSI C-ITS Security Standards

The European Cooperative Intelligent Transport Systems (C-ITS) architecture defines its security framework through ETSI TS 103 097 (security header and certificate formats) [8] and ETSI TS 102 941 (trust and privacy management) [54]. Similar to the IEEE approach, it employs an elliptic curve PKI with ECDSA signatures on 256-bit curves (typically NIST P-256, and optionally, Brainpool curves for specific applications) [80], [81]. Certificates in ETSI C-ITS are encoded using the Octet Encoding Rules (OER) rather than the traditional X.509 Distinguished Encoding Rules (DER) format, reducing overhead while preserving critical metadata, such as the issuer, validity period, and public key [82]. The trust model is layered, comprising entities, such as Root CA, Enrollment Authority (EA), and Authorization Authority (AA), in a cooperative PKI hierarchy [82]. Vehicles initially obtain a long-term enrollment certificate from an EA and subsequently receive periodic batches of short-lived authorization tickets (pseudonym certificates) from an AA. These pseudonym certificates are used to sign Cooperative Awareness Messages (CAMs), Decentralized Environmental Notification Messages (DENMs), and other V2X payloads [92]. The overall cryptographic framework in ETSI C-ITS standards balances strong authentication with vehicle identity privacy, relying on a multi-tier ECC-based certificate infrastructure for all cryptographic operations. Since ETSI C-ITS security profiles and OER certificate encodings are engineered for ECC-sized keys and signatures, similar to IEEE 1609.2, native PQC artifacts would significantly increase message and certificate sizes under current profiles.

### 4.3 SAE J2735 (Message Set Dictionary)

SAE J2735 defines the data structure and syntax for V2X messages, such as BSMs, SPaT messages, and Map Data (MAP) [83]. While it establishes the format and semantics of these messages, SAE J2735 does not define security mechanisms directly. Instead, the security of messages constructed under SAE J2735 is delegated to external standards, such as IEEE 1609.2 or ETSI TS 103 097, which wrap SAE J2735-formatted payloads with cryptographic protections [83]. For example, in the DSRC/WAVE stack, IEEE 1609.2 provides the certificate-based signing and verification layers around the application-level SAE J2735 content [7]. Thus, although SAE J2735 itself is algorithm-independent, its message structures are commonly transported within security envelopes defined by IEEE 1609.2 or ETSI C-ITS. Consequently, any PQC-driven growth in certificate and signature sizes propagates into the transport of J2735 messages, creating pressure for larger payload handling and more flexible length encoding in secured deployments.

### 4.4 3GPP (Cellular V2X) Standards

The Third Generation Partnership Project (3GPP) has defined V2X communication using cellular technologies in releases, such as Rel. 14/15 LTE-V2X and Rel. 16+ 5G NR V2X [84,85]. These specifications primarily focus on the radio interface and system architecture, rather than cryptographic security. For sidelink (PC5) direct peer-to-peer (P2P) communication between vehicles or between vehicles and infrastructure, 3GPP does not implement built-in link-layer encryption or authentication; instead, it relies on upper-layer security provided by established PKI frameworks, such as IEEE 1609.2 or ETSI C-ITS [86]. In practice, cellular V2X (C-V2X) messages are signed using pseudonym certificates provisioned through Security Credential Management Systems (SCMS) or the European Union C-ITS infrastructure. Vehicles store these certificates in their OBUs and authenticate messages using standard ECC-based algorithms. For network-layer communications (e.g., V2C and V2N), SIM-based authentication under 3GPP is used and is distinct from V2X message security layers. The key point is that 3GPP V2X leverages existing certificate infrastructures instead of introducing new cryptographic primitives. As a result, PQC readiness in C-V2X depends directly on the evolution of IEEE 1609.2 and ETSI C-ITS certificate and signature profiles, and PQC migration will be introduced through hybrid and post-quantum credential frameworks layered on top of the 3GPP communication protocols.

### 4.5 ISO/SAE 21434 (Automotive Cybersecurity)

ISO/SAE 21434:2021 is a comprehensive, process-oriented standard for managing cybersecurity throughout the automotive lifecycle [87]. It emphasizes risk-based decision-making and mandates that automakers and suppliers implement appropriate controls to mitigate cybersecurity threats. Although cryptography is identified as a fundamental control, the standard is algorithm-agnostic and does not mandate specific cryptographic techniques. Implementers are instead expected to use widely deployed and standardized cryptographic algorithms, such as AES, RSA, or ECC and ensure crypto-agility (defined as the ability to update or negotiate cryptographic algorithms in the field) [87]. Currently, ISO/SAE 21434 does not require PQC but acknowledges that cryptographic mechanisms must be kept up to date. Therefore, while the standard lays the foundational principles for secure automotive systems, the specific adoption of PQC remains at the discretion of OEMs and system designers.

### 4.6 UNECE WP.29 Regulations (UN R155/R156)

The United Nations Economic Commission for Europe (UNECE) has introduced two regulatory frameworks relevant to vehicular cybersecurity: UN R155 on Cybersecurity [88] and UN R156 on Software Updates [89]. These regulations have been widely adopted across global markets and oblige manufacturers to implement comprehensive cybersecurity management systems. UN R155 mandates that automakers identify and mitigate cyber risks throughout the vehicle lifecycle and requires that all cryptographic modules used in the vehicle comply with industry-recognized standards. While the regulations do not prescribe specific cryptographic algorithms, they enforce the use of "state-of-the-art" protections, implicitly requiring strong encryption and digital signatures for secure boot, diagnostics, and V2X communication. Many manufacturers have implemented these protections using ECC-based certificate infrastructures, such as X.509 or IEEE 1609.2-compliant formats [90]. Although WP.29 (World Forum for Harmonization of Vehicle Regulations) does not currently explicitly reference PQC, its emphasis on evolving best practices suggests that future versions may require the adoption of PQC algorithms once they are standardized and validated. Regulatory lag remains a challenge. Under the current UNECE WP.29 framework, manufacturers aiming to deploy PQC-enhanced systems are required to provide detailed migration plans, performance benchmarks, and compatibility evidence to obtain regulatory approval [88], [89].

### 4.7 AUTOSAR

The Automotive Open System Architecture (AUTOSAR) defines a modular cryptographic framework for vehicle electronic control units (ECUs) through its Adaptive Platform Crypto Stack and related security modules such as Security Onboard Communication (SecOC) [91,92]. The AUTOSAR Adaptive Cryptography specifications define standardized crypto-provider interfaces, secure key-slot management, algorithm identifiers, and protected-key handling mechanisms, including support for KEM abstractions and hardware security module (HSM)-backed implementations [91,93]. Notably, the standard intentionally avoids locking in a fixed set of algorithms and instead allows provider-defined algorithm identifiers, enabling future extensibility and provider-specific algorithm support [93]. Research extensions to SecOC, such as the Confidential, Integral, aNd Authentic on board coMunicatiON (CINNAMON) module, further illustrate ongoing efforts to enhance secure onboard communication within the AUTOSAR ecosystem [94]. However, current AUTOSAR specifications and SecOC profiles do not yet define or standardize PQC algorithms or parameter profiles [91,93]. As a result, while the architecture is structurally crypto-agile and can accommodate new primitives such as PQC KEMs and signature schemes at the interface level, practical PQC deployment in AUTOSAR-based in-vehicle networks would require explicit standard profiles, provider support, and validation guidance [92,93].

To synthesize the key insights from subsections 4.1 to 4.7, **Table 9** compares current vehicular cryptographic standards with respect to their certificate structures, trust models, PQC readiness, and architectural scope. The comparison reveals substantial variability in how well current standards are positioned for a post-quantum transition, suggesting that coordinated updates will be needed to ensure secure and interoperable cryptographic migration across ITS environments.

**Table 9**
**Comparative overview of current vehicular communication and security standards and their readiness for PQC integration**

| Standard | Primary Cryptographic Methods | Certificate Format and Trust Model | PQC Support | Key Scope/Notes |
|---|---|---|---|---|
| IEEE 1609.2 | ECDSA (P-256), ECIES + AES-CCM | Implicit/explicit certificates; Hierarchical PKI (Root CA →Sub-CAs) | No | Optimized for DSRC bandwidth; short-lived pseudonyms for privacy |
| ETSI C-ITS | ECDSA (P-256/Brainpool) | OER-encoded pseudonym certificates; Layered PKI (Root CA, EA, AA) | No | Balances authentication with privacy; used for CAMs/DENMs in EU |
| SAE J2735 | Delegated to IEEE 1609.2 or ETSI | N/A (Message format standard only) | No | Defines V2X message (BSMs, SPaT); relies on external security wrappers |
| 3GPP (C-V2X) | ECDSA (via IEEE/ETSI PKI) | Pseudonym certificates (SCMS/C-ITS); SIM authentication for network communications | No | Focuses on cellular architecture; defers link-layer security to IEEE/ETSI PKI |
| ISO/SAE 21434 | Agnostic (AES, RSA, ECC recommended) | N/A (Process standard) | Indirect | Mandates crypto-agility but no PQC requirement; risk-based lifecycle management |
| UNECE WP.29 (R155) | Agnostic ("state-of-the-art" ECC typical) | X.509/IEEE 1609.2-compliant PKI | Indirect | Requires compliance with industry standards; expects migration to future PQC |
| AUTOSAR | Symmetric + Asymmetric (provider-defined; e.g., MAC, ECC, KEM interfaces) | Not standardized; depends on external PKI or application profile | No | Secures ECU-to-ECU communication; crypto-provider architecture is extensible, but PQC algorithms are not yet standardized |

Notes- AA: Authorization Authority; AES-CCM: Advanced Encryption Standard-Counter with CBC-MAC; CA: Certificate Authority; CAMs: Cooperative Awareness Messages; C-V2X: Cellular-V2X; DENMs: Decentralized Environmental Notification Messages; DSRC: Dedicated Short-Range Communication; EA: Enrollment Authority; EU: European Union; ECC: Elliptic Curve Cryptography; ECDSA: Elliptic Curve Digital Signature Algorithm; ECIES: Elliptic Curve Integrated Encryption Scheme; ECU: Electronic Control Unit; KEM: Key Encapsulation Mechanism; MAC: Message Authentication Code; N/A: Not Applicable; OER: Octet Encoding Rules; PKI: Public Key Infrastructure; SCMS: Security Credential Management System; SIM: Subscriber Identity Module; SPaT: Signal Phase and Timing; V2X: Vehicle-to-Everything.

## 4.8 Research Gaps in Standardization and Certification Frameworks for Post-Quantum Vehicular and Transportation Infrastructure Security

As summarized in **Fig. 3**, the research gaps in PQC-ready standardization cluster into three primary domains, i.e., regulatory frameworks, vehicular cybersecurity, and V2X communication security. These domains group together standards, such as IEEE 1609.2, ETSI C-ITS, ISO/SAE 21434, UNECE WP.29, and AUTOSAR, based on their functional focus within the ITS security ecosystem. The figure also shows how these domains intersect through shared dependencies on ECC-based PKI infrastructures and overlapping compliance requirements. The research gaps identified in this section were mapped to these domains, and they are presented below:

- **Research Gap 1** (**RG1**) concerns the lack of quantum-safe certificate profiles in standards, such as IEEE 1609.2 and ETSI C-ITS, which currently do not support embedding large PQC public keys (e.g., ML-DSA [Dilithium] keys) within compact formats like Implicit Certificates. These limitations complicate the deployment of PQC signatures within existing certificate structures without exceeding message size budgets.
- **RG2** arises from regulatory ambiguity. Standards, such as UN R155 and ISO/SAE 21434, mandate state-of-the-art cryptography but provide no concrete timelines, benchmarks, or validation criteria for PQC deployment in safety-critical applications. This lack of specificity hinders migration planning and undermines industry-wide consistency.

- **RG3** reflects the absence of standardized cross-certification frameworks between classical ECC-based and PQC-based trust anchors. Without such mechanisms, different vehicles and infrastructure components may trust different certificate authorities, creating gaps in mutual authentication and disrupting interoperability.

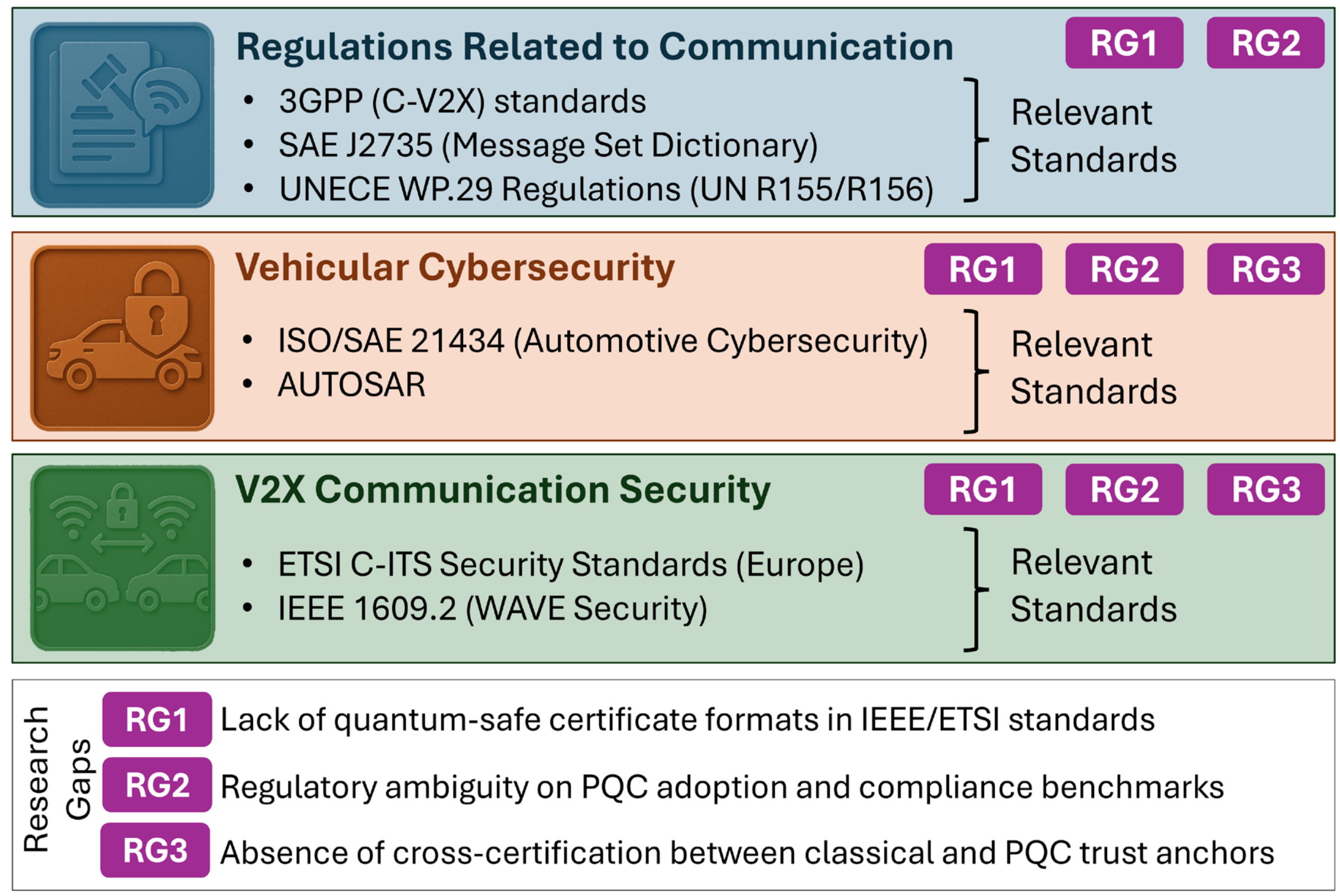

**Fig. 3.** Research gaps (RG1-RG3) mapped to PQC-ready standardization and certification frameworks across regulatory, vehicular cybersecurity, and V2X communication domains.

### 4.9 Concluding Remarks of Section 4

Current vehicular communication and cybersecurity standards are strongly dependent on ECC-based PKI structures and certificate formats that are not yet PQC-ready at the artifact and profile level. The cross-standard comparison highlights a consistent pattern: several frameworks endorse crypto agility in principle but lack explicit PQC certificate profiles, migration benchmarks, and cross-certification mechanisms needed for interoperable deployment. The synthesis across IEEE, ETSI, SAE, 3GPP, ISO/SAE, UNECE, and AUTOSAR reveals that PQC migration primarily involves updating certificate formats, protocol structures, and interoperability frameworks, rather than simply replacing cryptographic algorithms. Nevertheless, this review is limited to publicly available standards texts and technical reports and does not include unpublished industry migration plans or proprietary SCMS redesign efforts. Regulatory interpretation and certification workflows were also analyzed at a technical level only, without jurisdiction-specific legal analysis. Future updates to standard drafts and pilot deployments may therefore change the readiness landscape faster than reflected here. The next section examines how PQC can be practically integrated across in-vehicle, OEM, and V2X communication interfaces, considering performance trade-offs, protocol constraints, and deployment feasibility.

## 5. POST-QUANTUM CRYPTOGRAPHY FOR ITS COMMUNICATION INTERFACES

Modern vehicles increasingly rely on secure communication among vehicle subsystems (in-vehicle networks), with the OEM/cloud backend channels, and with other vehicles and transportation infrastructures (V2X), including V2V, V2I, V2P, and I2I. Classical schemes, such as ECIES and ECDSA, form the basis of today's automotive PKI [31] and are known to be vulnerable to quantum algorithms such as Shor's algorithm. Consequently, deploying PQC offers long-term cryptographic resilience but introduces new performance and integration challenges in bandwidth-constrained, latency-sensitive vehicular environments. This section discusses the current state of PQC adoption in these communication interfaces and presents empirical benchmarking results, technical constraints, and deployment strategies supported by the literature.

### 5.1 In-Vehicle Network Security

#### *5.1.1 Legacy Systems and Cryptographic Gaps*

In-vehicle networks form the communication backbone for coordinating safety-critical and operational functions among ECUs. Protocols, such as CAN [95], FlexRay [96], Local Interconnect Network (LIN) [97], Media Oriented Systems Transport (MOST) [98], and Automotive Ethernet [99], facilitate intra-vehicular communication across a range of applications, including braking, steering, diagnostics, and infotainment [100]. However, these legacy protocols were designed without integrated cryptographic features. In particular, the classic CAN protocol lacks built-in support for authentication or encryption mechanisms, making it vulnerable to a wide range of cybersecurity threats, including spoofing, message injection, replay attacks, and denial-of-service (DoS) attacks [101–103]. As vehicles become increasingly connected and software-driven, these limitations expose security gaps that require modern cryptographic protections, including PQC algorithms.

Although quantum adversaries may not directly exploit CAN, FlexRay, or Automotive Ethernet traffic in real time, the more realistic threat lies in 'harvest-now-decrypt-later' attacks on logged vehicular data or on OTA update traffic that traverses insecure in-vehicle channels before reaching secure backhaul links. Furthermore, as in-vehicle networks increasingly connect to external V2X or cloud domains, a quantum-safe baseline becomes necessary to ensure end-to-end protection and to prevent legacy in-vehicle protocols from becoming the weakest link in a secure ITS ecosystem.

#### *5.1.2 PQC Integration Approaches by Protocol*

To mitigate the cryptographic deficiencies of legacy in-vehicle networks, researchers have explored the integration of PQC into existing in-vehicle communication protocols under strict payload, timing, and hardware constraints. Because most PQC schemes generate substantially larger keys, ciphertexts, and signatures than classical ECC-based mechanisms, protocol-specific integration strategies, especially hybrid designs that combine PQC algorithms with classical cryptographic mechanisms, are commonly used to balance security and feasibility.

In CAN-FD environments, PQC is typically introduced through hybrid constructions that separate key establishment and data protection. CAN-FD extends the original CAN payload from 8 bytes to 64 bytes and supports data rates up to 8 Mb/s [104], [105], enabling limited carriage of cryptographic metadata. Within this constrained setting, lattice-based KEMs such as CRYSTALS-Kyber are commonly used to establish a shared session key, after which symmetric encryption (e.g., AES, RC6, or Twofish) protects application data [101]. Hardware-enhanced approaches combining Physically Unclonable Functions (PUFs) with PQC key exchange mechanisms (e.g., SIDH or Kyber) have also been proposed to strengthen resistance against device-level attacks [106]. However, in these designs, PQC artifacts are not transported as a single Secure Protocol Data Unit (SPDU), but are distributed across multiple frames, requiring explicit fragmentation and reassembly logic at higher protocol layers [101].

FlexRay, which is used in deterministic, time-critical subsystems such as braking and powertrain control, presents a different constraint profile. Its time-triggered slot structure supports predictable latency but limits payload flexibility. Although lightweight lattice-based PQC variants can meet computational latency targets [16,103], FlexRay's fixed communication schedule and limited slot payload (maximum 254 bytes) make large PQC artifacts difficult to transport without redesigning the cycle-level schedule, with fragmentation handled only through higher-layer segmentation mechanisms. As a result, most PQC proposals for FlexRay focus on hybrid or gateway-assisted designs rather than full native PQC message embedding.

Automotive Ethernet provides a substantially more favorable platform for PQC deployment due to its higher bandwidth and compatibility with secure communication stacks, such as TLS and MACsec [99]. PQC algorithms, including NTRUEncrypt and CRYSTALS-Dilithium, have been integrated into Ethernet-based secure boot, authentication, and update pipelines [102,107–111]. Compared to other in-vehicle communication protocols, Ethernet's higher throughput and larger frame size (approximately 1,500 bytes) allow PQC keys and signatures to be transported with reduced fragmentation pressure. This makes Ethernet-based in-vehicle segments the primary early candidates for full-PQC authentication and key-establishment workflows. The availability of multicore automotive processors and HSMs, including designs assisted by Field-Programmable Gate Array (FPGA) and Application-Specific Integrated Circuit (ASIC) technologoes, further improves PQC feasibility by accelerating polynomial and hashing operations under real-time constraints [16,31].

#### *5.1.3 Benchmark Analysis*

Benchmarking studies have revealed specific limitations in executing PQC algorithms on automotive-grade processors. However, reported latency figures across the literature are often measured on different processor architectures and clock speeds, making direct time-based comparisons misleading unless normalized. Accordingly, in addition to the latency, the results below are interpreted using the operation class and relative cost trends.

On the computational side, Dilithium verification requires approximately 2 times as many CPU cycles as the Edwards-curve Digital Signature Algorithm (EdDSA) on similar hardware [112]. Hash-based XMSS (eXtended Merkle Signature Scheme) PQC signature also exhibits long delays, with verification times ranging from 194 to 518 milliseconds (ms) on a Cortex-M4 [112]. When mapped to cycle counts using typical Cortex-M4 clock ranges (100-200 MHz), these delays correspond to tens of millions of cycles per verification, confirming that hash-based signatures are generally unsuitable for tight real-time control loops on resource-constrained microcontroller units (MCUs) without hardware acceleration. Consistent MCU-level measurements on automotive controllers, such as Infineon AURIX platforms, also show that lattice-based KEM operations (e.g., Kyber-512) remain feasible within millisecond-scale budgets, whereas heavier constructions incur significantly higher cycle costs [17,113].

In terms of memory footprint, Falcon signatures (1.28 kilobytes [KB]) exceed the capacity of MCUs at higher security levels [110,114], and Dilithium keys (1.3-2.6 KB) demand an order of magnitude more storage than ECDSA keys [110,115]. This storage expansion directly affects secure boot key storage, certificate caching, and ECU flash layout in embedded automotive controllers.

Operational asymmetry presents an additional challenge. For example, McEliece decryption is reported to be approximately six orders of magnitude slower than encryption in software simulation [116], making hardware acceleration essential for practical implementation. Such asymmetry is particularly problematic for receiver-heavy workflows (e.g., signature verification, KEM decapsulation) common in broadcast authentication and secure update validation.

To improve comparability, **Table 10** groups benchmarking results by platform class (MCU, automotive system-on-chip [SoC], FPGA/HSM) and reports the original measured latency along with the clock speed. Readers should interpret performance trends within each platform class rather than comparing raw milliseconds across different clock domains. Because reported benchmarks are obtained on processors with different clock frequencies and microarchitectures, direct latency comparison is not meaningful. A rough normalization can be made using the clock rate (cycles ≈ latency × clock frequency), but actual performance also depends on the memory hierarchy, instruction set features, and the presence of hardware accelerators. Therefore, normalized cycle estimates in this table are first-order approximations only, and conclusions are drawn primarily from intra-platform comparisons.

**Table 10**
**Representative performance metrics and implementation constraints of PQC algorithms for in-vehicle networks (with approximate cycle normalization)**

| Ref. | Platform Class | Device | Clock Speed | PQC Algorithm | Operation | Performance (Latency/ Throughput) | Approximate Cycles (Normalized) | Comment |
|---|---|---|---|---|---|---|---|---|
| [101] | Simulation | Linux VM (Socket CAN) | N/A | Kyber-512 | CAN simulation | <0.5% latency overhead | N/A | Bandwidth limits in CAN |
| [116] | Automotive SoC | ARM Cortex-A53 | 1.1 GHz | McEliece (block size, k = 2056) | SW encryption | 231.79 ms | $\approx 2.55\times10^{8}$ | Software encryption reasonable |
| [116] | Automotive SoC | ARM Cortex-A53 | 1.1 GHz | McEliece (block size, k = 2056) | SW decryption | $1.62\times10^{8}$ ms | $\approx 1.78\times10^{14}$ | Software decryption impractical |
| [111] | Automotive SoC | NXP S32G274A | 800 MHz | NTRUEncrypt | Encryption | 0.28 ms | $\approx 2.24\times10^{5}$ | 66 times faster than ECDH |
| [108] | MCU | TMS570 LS3157 (Cortex-R4) | 160 MHz | Kyber + Dilithium | Full authentication | 1.676 s (42 times bandwidth increase) | $\approx 2.68\times10^{8}$ | CAN-FD fragmentation required |

| Ref. | Platform Class | Device | Clock Speed | PQC Algorithm | Operation | Performance (Latency/ Throughput) | Approximate Cycles (Normalized) | Comment |
|---|---|---|---|---|---|---|---|---|
| [112] | MCU | ARM Cortex-M4 | 168 MHz | Dilithium | Signature verification | 37 ms (2 times slower than EdDSA) | $\approx 6.22\times10^6$ | Memory-intensive (1.5 KB keys) |
| [112] | MCU | ARM Cortex-M4 | 168 MHz | XMSS | Signature verification | 518 ms (XMSS C); 194 ms (XMSS ASM) | $\approx 87.12\times10^6$<br>$\approx 32.64\times10^6$ | Fails real-time MCU verification budgets |
| [115] | Automotive SoC | NXP S32G274A | 400 MHz | Dilithium2 | Secure boot verify | 11.3 ms (w/SHA-256 pre-hash) | $\approx 4.52\times10^6$ | Boot time optimization |
| [117] | Automotive SoC | AURIX TC397 (6-core) | 300 MHz | NTRUEncrypt | Multicore keygen | 43% faster vs single-core | N/A | Parallelizes polynomial inversions |
| [112] | FPGA / HSM | Artix-7 FPGA | 123-269 MHz | Kyber/ BIKE | HSM operations | 0-13% Look-Up Table (LUT) overhead | N/A | Area-efficient co-design |
| [17] | MCU | AURIX TC297TF | 300 MHz | Kyber-512 | Full KEM operations | ≈ 8.18 ms | $2.45\times10^6$ | MCU-feasible lattice KEM |
| [17] | MCU | AURIX TC297TF | 300 MHz | Saber | Full KEM operations | ≈ 21.18 ms | $6.35\times10^6$ | Higher cost vs. Kyber |
| [113] | MCU | AURIX TC397x | 300 MHz | Kyber-512 (optimized) | Full KEM operations | ≈ 8.5 ms total | $2.54\times10^6$ | Minimal memory footprint (18.9 KB) |

Notes- Approximate Cycles (Normalized): estimated cycle count computed from reported latency and device clock speed (cycles ≈ latency × clock frequency) or from throughput where available; CAN-FD: Controller Area Network-Flexible Data Rate; ECC: Elliptic Curve Cryptography; ECDH: Elliptic Curve Diffie-Hellman; EdDSA: Edwards-curve Digital Signature Algorithm; FPGA: Field-Programmable Gate Array; HSM: Hardware Security Module; KEM: Key Encapsulation Mechanism; MCU: Microcontroller Unit; N/A: not available; Ref.: reference number; SHA: Secure Hash Algorithm; SoC: System-on-chip; SW: software.

#### *5.1.4 Challenges for PQC Migration in In-Vehicle Networks*

A major bottleneck for PQC deployment in in-vehicle networks lies in the hardware capabilities of embedded ECUs. Most automotive ECUs operate within tight hardware constraints, typically featuring low clock frequencies (10-300 MHz) and limited memory (40-100 KB) [4,16,103,109]. These constraints directly limit the feasibility of compute- and memory-intensive PQC KEM and signature schemes on safety-critical control units.

Real-time timing requirements further restrict adoption. Safety-critical functions, such as steering and braking, often require cryptographic processing within 10 ms [16,102,103]. Measured results show that Dilithium secure-boot verification can take about 11.3 ms compared to 2.7 ms for RSA on 400 MHz NXP S32G274A [115], exceeding typical boot-stage timing budgets. Even when individual PQC operations are fast on higher-end processors (e.g., NTRUEncrypt ≈0.28 ms on 800 MHz [102,111]), system-level delay increases when larger PQC artifacts trigger packet fragmentation and reassembly. Multicore parallelization (e.g., AURIX TC397) can reduce key-generation latency by approximately 43% [117], but adds complexity to certification and fault tolerance.

Network payload limits create additional measurable overhead. CAN (8 bytes) and CAN-FD (64 bytes) frames cannot carry PQC artifacts without fragmentation. A Kyber-512 ciphertext (≈768 bytes) requires at least 12 CAN-FD frames, versus typically 1-2 frames for ECC artifacts [101,106]. FlexRay payload limits (≈254 bytes) similarly force multi-frame transmission and constrained scheduling [16,103]. Although MOST150 supports 384-byte frames and LIN allows up to 8 bytes, their roles are typically limited to infotainment or low-priority functions, making them unsuitable for time-critical cryptographic operations. The industry migration toward Ethernet (≈1,500 bytes) for bandwidth-intensive applications reflects these limitations. Bandwidth overhead for PQC can be substantial. For example, integrating Kyber and Dilithium into Lightweight Authentication for Secure Automotive Networks (LASAN_M) authentication increases bandwidth use by 42 times relative to classical cryptographic protocols [108].

Crypto-agility is also limited in current platforms. Many in-vehicle accelerators are fixed-function (AES, SHA-2), making PQC upgrades difficult without hardware changes [102,116]. Additionally, common standards such as AUTOSAR SecOC lack built-in mechanisms for cipher negotiation or migration [92,102]. Existing regulations (ISO/SAE 21434, UN R155) require long-term cryptographic robustness but do not yet define PQC migration procedures for in-vehicle networks [16,88,107]. Accordingly, practical migration approaches focus on programmable HSMs [115,120,121], hybrid classical-PQC operation [106], protocol optimizations such as CAN-FD key pre-distribution [101,106], and phased ECU upgrades for safety-critical modules [16,17].

## 5.2 OEM-Specific Communications

OEM-specific communications include operations, such as firmware and software updates, remote diagnostics, feature activation, and secure V2C exchanges, managed within the OEM ecosystem. Without strong cryptographic protections, attackers could exploit these interfaces to inject malicious firmware, manipulate diagnostic data, or remotely access and control vehicle functionalities, potentially leading to catastrophic safety, privacy, and operational consequences. As vehicles transition into highly connected, software-defined platforms, the resilience of these channels against both classical and quantum adversaries becomes vital. OEMs face unique challenges due to long vehicle lifecycles, heterogeneous embedded hardware across ECUs, and evolving regulatory mandates, such as UN R155 and ISO/SAE 21434 [109,118]. Accordingly, PQC adoption in OEM channels is driven not only by cryptographic strength but also by boot-time budgets, update latency tolerance, certificate size, and the availability of hardware support.

### *5.2.1 Secure Firmware Updates and Boot Verification*

Secure boot and firmware update workflows depend on digital signature verification to guarantee authenticity and integrity. Classical RSA and ECC signatures are widely used today but are not quantum-resistant. Consequently, PQC signature schemes, including Leighton-Micali Signature (LMS) [119], SPHINCS+, Dilithium, and Falcon, have been evaluated for secure boot and OTA validation pipeline [120–122].

Hash-based LMS demonstrates strong suitability for embedded boot verification due to low verification latency and moderate signature size. Measurements on Intel Xeon CPU (2.20 GHz) show verification latency between 0.17-1.30 ms with LSM signatures between approximately 1.6-2.8 KB, and FPGA implementations (in Xilinx UltraScale chips) report sub-10 ms verification using modest logic resources [120]. Variants of SPHINCS+ provide stateless security but with larger signatures (≈ 8-23 KB) and verification latency between 0.47-6.97 ms on an Intel Xeon CPU (2.20 GHz) platform [120]. This places SPHINCS+ near or beyond typical secure-boot timing budgets for many low-end ECUs when scaled from server-class measurements to embedded clock domains, thereby limiting its practicality for early-stage boot verification without hardware acceleration or parameter tuning.

Lattice-based signatures offer an alternative tradeoff between performance and cryptographic artifact size. Dilithium verification has been demonstrated at 0.5 ms on Cortex-A53 (1 GHz) and 11.1 ms on Cortex-M7 hardware security engine (HSE) platforms [122], making it feasible for secure boot on higher-end automotive processors. Falcon achieves smaller signatures (≈ 0.7-1.3 KB), but verification latency increases with firmware image size due to SHAKE-256 hashing, reaching up to 118 ms for 5.9 MiB (1 mebibyte [MiB] = $2^{20}$ bytes) images on Xilinx ZCU102 platforms, compared to 28 ms for RSA [121]. This size-dependent verification cost makes Falcon attractive for bandwidth-constrained certificate and credential transport, but introduces image-size-dependent latency for large OTA package validation, which must be accounted for in boot and update timing budgets.

Secure boot verification typically requires sub-10 ms execution to avoid startup delay [16,102,103], whereas OTA updates can tolerate higher verification latency because they execute outside real-time control paths. Reported optimization strategies include SHAKE-256 hardware acceleration, floating-point support for Falcon arithmetic, hierarchical signature trees for revocation compatibility, pre-hashing of large firmware images, and tamper-protected public-key storage [120–122]. HSEs further improve fault resistance but can introduce modest verification overhead (≈11 ms for Dilithium-2 on Cortex-M7 HSE), which is generally acceptable in OTA workflows, as demonstrated in AUTOSAR Update and Configuration Management (UCM) implementations [122].

### *5.2.2 Secure Diagnostics and Telematics Communication*

OEM telematics and diagnostics channels are increasingly being evaluated using PQC-enabled Transport Layer Security (TLS) 1.3 and Quick UDP Internet Connections (QUIC) handshakes for V2C communication [123,124]. Benchmarks on automotive-class processors show that Dilithium achieves high signing throughput (≈881 ops/sec on Cortex-A72 at 1.5 GHz), while Falcon achieves lower signing throughput (≈77 ops/sec) on the same device but benefits from smaller certificate sizes (≈0.7-1.3 KB) [123]. As summarized in **Table 11**, these measurements illustrate trade-off between cryptographic computation cost and certificate transmission overhead on constrained telematics links.

Despite larger certificate sizes, QUIC combined with Dilithium has been shown to outperform TCP/TLS 1.3 with ECDSA in terms of session establishment time under lossy network conditions, with experiments reporting up to 73.2% faster handshake/session completion at 5% packet loss [124]. This suggests that transport-layer behavior can partially offset the cryptographic overhead of PQC under realistic network impairments, particularly for long-lived or loss-prone telematics sessions. In addition, anonymous lattice-based authentication protocols such as RAVEN demonstrate low authentication latency (≈3.9 ms) and reduced communication overhead (≈1,201 bits) compared with ECC-based approaches when evaluated on a high-end processor platform (≈2.2 GHz CPU) [125], indicating feasibility for privacy-preserving V2C authentication, although embedded automotive processors would be expected to show higher latency without hardware acceleration.

Overall, OEM communication channels tolerate PQC overhead more readily than in-vehicle control networks because staged secure-boot verification and telematics sessions operate under comparatively more flexible timing budgets than real-time control and safety-critical message loops. In addition, these OEM and backend-facing functions are more likely to run on higher-performance processors or dedicated security modules. **Table 11** further reports normalized cycle estimates to support intra-platform comparison across PQC signature options.

**Table 11**
**Selected benchmarking results of PQC signature and authentication schemes for OEM-specific automotive communications**

| Ref. | Platform | Device (Clock Speed) | PQC Algorithm | Operation | Performance (Latency/ Throughput) | Signature Size | Approx. Cycles (Normalized) |
|---|---|---|---|---|---|---|---|
| [120] | Server CPU | Intel Xeon (2.20 GHz) | LMS (hash) | Verification | 0.17-1.30 ms | 1.8-5.8 KB | $\approx 3.7\times10^{6}$ - $2.9\times10^{6}$ |
| [120] | Server CPU | Intel Xeon (2.20 GHz) | SPHINCS+ | Verification | 0.47-6.97 ms | 15-23 KB | $\approx 1.0\times10^{6}$ - $1.53\times10^{7}$ |
| [122] | Automotive SoC | NXP S32G (Cortex-A53) (1 GHz) | Dilithium | OTA Verification | 0.5 ms | 2,420 bytes | $\approx 5.0\times10^{5}$ |
| [122] | Automotive SoC + HSE | NXP S32G (Cortex-M7 HSE) (400 MHz) | Dilithium | OTA Verification | 11.1 ms | 2,420 bytes | $\approx 4.4\times10^{6}$ |
| [121] | Automotive SoC | Xilinx ZCU102 (Cortex-A53 quad) | Falcon | OTA Verification | 22-118 ms (1.1-5.9 MiB file) | 0.9-1.3 KB | N/A |
| [121] | Automotive SoC | Xilinx ZCU102 (Cortex-A53 quad) | Dilithium | OTA Verification | 28-139 ms (1.1-5.9 MiB file) | 2.5-4.5 KB | N/A |
| [123] | Edge SoC | ARM Cortex-A72 (RPi 4) (1.5 GHz) | Falcon | Sign | 76.7 ops/sec | 0.7-1.3 KB | $\approx 2.0\times10^{7}$ cycles/op |
| [123] | Edge SoC | ARM Cortex-A72 (RPi 4) (1.5 GHz) | Dilithium | Sign | 881.8 ops/sec | 2.4-7.5 KB | $\approx 1.7\times10^{6}$ cycles/op |
| [125] | CPU Simulation | Intel i7-5700HQ (simulated) 2.8 GHz | RAVEN (Lattice) | Authentication | 3.885 ms | 1,201 bits | $\approx 1.09\times10^{7}$ |

Notes- Approx. Cycles (Normalized): estimated cycle count derived from reported latency or throughput and clock speed (cycles ≈ latency × clock frequency, or clock frequency ÷ ops/sec); HSE: Hardware Security Engine; KB: kilobytes; MiB: mebibytes (1 MiB = $2^{20}$ bytes); N/A: not available; ops/sec: operations per second; OTA: Over-the-Air (update); Ref.: reference number; SoC: System-on-chip.

## 5.3 V2X Communication Security

V2X communication enables the real-time exchange of data among vehicles, infrastructure, pedestrians, and cloud services, forming the basis for safety-critical applications such as collision warnings, cooperative driving, and dynamic traffic management. Current frameworks, including IEEE 1609.2 and ETSI C-ITS, rely on classical cryptography, such as ECDSA and ECIES, which are vulnerable to Shor's algorithm. Ensuring long-term confidentiality and trust in V2X ecosystems thus requires transitioning to quantum-resistant cryptographic mechanisms. V2X security objectives, i.e., authentication, integrity, confidentiality, non-repudiation, and availability, must be achieved within strict timing constraints of ≤100 ms for safety-critical BSM authentication and <5 ms for reliable functions such as emergency braking [83]. Accordingly, practical deployments require PQC selections and integration strategies that preserve security while limiting computational and communication overhead.

### *5.3.1 V2X PQC Application Scenarios: Requirements and Security Objectives*

V2X security spans multiple interaction modes (V2V, V2I, V2N, V2P), each with different resource constraints and trust requirements. I2I communication is also commonly treated as part of the V2X ecosystem, in which RSUs, traffic signals, and transportation infrastructure exchange authenticated data to coordinate operations and support safe mobility.

In V2V, periodic broadcast safety messaging (e.g., BSMs at 10 Hz, i.e., every 100 ms) requires low-latency authentication [83]. Lattice-based signatures such as Falcon-512 (NIST security level 1) have demonstrated feasibility in this setting, particularly in hybrid designs that reduce per-message overhead. For example, when Falcon-512 is combined with classical ECDSA (P-256) in a partially hybrid design, where Falcon protects credential material during certificate provisioning and ECDSA signs routine broadcast messages, channel load is substantially lower than with full-PQC signing, and the scheme remains feasible within DSRC payload constraints, whereas pure PQC signing is shown to be impractical for V2V under current standards [126]. System-level V2V studies further show that PQC feasibility depends strongly on certificate size, fragmentation behavior, and transmission scheduling, and that hybrid or staged credential strategies are more channel-efficient than per-message full-PQC signing [127].

In V2I scenarios (e.g., tolling, traffic signal coordination, and certificate provisioning), secure session establishment typically has more tolerant bandwidth and latency budgets than broadcast authentication. Experimental measurements of post-quantum signatures used within TLS 1.3 show that Dilithium-2 signing requires 0.41-0.82 ms and verification 0.12-0.16 ms, while Falcon-512 signing requires 5.2-6.5 ms and verification 0.05-0.07 ms on local and cloud Intel Xeon-class platforms. For comparison, ECDSA-384 signing and verification require about 1.3 ms and 0.9-1.0 ms, respectively, on the same platforms [128]. Falcon-512 further reduces bandwidth pressure with signatures of approximately 666 bytes, which fit within a single DSRC / ETSI ITS-G5 maximum transmission unit (MTU) (≈1492 bytes) without fragmentation, making it attractive for constrained V2X links [30].

In V2N and V2P contexts (e.g., smartphone-cloud interactions), preserving data privacy becomes dominant. Lattice-based homomorphic encryption (HE) can support privacy-preserving analytics for I2I and V2C by enabling computation over encrypted data. Experiments using post-quantum HE schemes, including BFV, BGV, and CKKS, report end-to-end latencies of 3-32 seconds in an RSU-Cloud pipeline, which are feasible for non-real-time ITS services while preserving confidentiality [73].

Provisioning and trust-routing studies also indicate that PQC-based credential and key-establishment workflows can remain scalable at the network level. For example, PQC-enabled SCMS provisioning and lattice-based V2I trust-routing simulations report sub-5 ms key exchange latency and under-30 ms routing/authentication latency at 500 nodes, showing that lattice-based designs can scale when certificate distribution and routing logic are protocol-aware [129].

Finally, some PQC families remain impractical for V2X. For example, multivariate schemes such as Rainbow have been reported to produce oversized certificates (e.g., 2,300 KB), causing handshake failures even on high-end platforms [4,128].

### *5.3.2 Computational Performance of Edge Devices*

Efficient execution of PQC on edge platforms, including OBUs and RSUs, is essential for practical V2X deployment. Reported benchmarks across cooperative ITS platforms show that lattice-based schemes (Falcon, Dilithium, Kyber) provide the most favorable security-latency balance under V2X timing constraints, while hash-based signatures such as SPHINCS+ exhibit substantially higher computational and bandwidth cost [30,130].

Cross-platform cooperative ITS benchmarking over representative edge device classes, for example, Low-End (LE) OBU 2022 (ARM Cortex-A53 class), High-End (HE) OBU 2024 (ARM Cortex-A72 class), RSU 2015 (Intel x86 class), and Server PKI 2021 (AMD x86 class), shows that Falcon-512 and Dilithium-2 achieve higher verification throughput (thousands of verifications per second) than ECDSA P-256 on higher-end OBUs and RSUs, whereas SPHINCS+ variants remain limited to a few hundred verifications per second, constraining their suitability for high-rate broadcast authentication [30].

Protocol-level experiments over the IEEE WAVE stack further confirm feasibility on modern OBU hardware. Measurements on an NVIDIA Jetson AGX Xavier platform (HE OBU) report total cryptographic processing times of 3.9 ms for ECDSA-P256 and 12.7 ms for Falcon (Level-1) across 300 WAVE payloads, remaining well within the 100 ms BSM authentication window despite Falcon's higher per-message cost [130]. Representative cross-platform PQC verification benchmarks and normalized cycle estimates for OBU and RSU classes are summarized in **Table 12**, including both measured throughput and clock-normalized cycle values to support within-class performance comparison.

**Table 12**
**Selected performance benchmarking of PQC algorithms for V2X edge devices (OBUs/RSUs)**

| Ref. | Platform Class | Device (Clock Speed) | Operation | Cryptographic Algorithm | Measured Performance (Latency/ Throughput) | Normalized Cycles / Verify (approx.) | Notes |
|---|---|---|---|---|---|---|---|
| [30] | LE OBU | ARMv8-A Cortex-A53 (1.2 GHz) | Verify | ECDSA P-256 | 1513 ops/s | $\approx 7.9\times10^5$ | Classical baseline; lower throughput than Falcon and Dilithium |
| | | | | Dilithium2 | 1721 ops/s | $\approx 7.0\times10^5$ | Exceeds ECDSA throughput |
| | | | | Falcon-512 | 5244 ops/s | $\approx 2.3\times10^5$ | Highest verification throughput in LE OBU class |
| | | | | SPHINCS+-128s | 201 ops/s | $\approx 6.0\times10^6$ | Hash-based; significantly slower |
| [30] | HE OBU | ARMv8-A Cortex-A72 (2 GHz) | Verify | ECDSA P-256 | 2895 ops/s | $\approx 6.9\times10^5$ | Higher-end OBU class |
| | | | | Dilithium2 | 4398 ops/s | $\approx 4.5\times10^5$ | PQC faster than ECDSA |
| | | | | Falcon-512 | 11155 ops/s | $\approx 1.8\times10^5$ | Best throughput |
| | | | | SPHINCS+-128s | 358 ops/s | $\approx 5.6\times10^6$ | Slowest class |
| [30] | RSU | Intel x86 (3.4 GHz) | Verify | ECDSA P-256 | 2763 ops/s | $\approx 1.2\times10^6$ | RSU class |
| | | | | Dilithium2 | 3219 ops/s | $\approx 1.1\times10^6$ | Comparable to ECDSA |
| | | | | Falcon-512 | 11215 ops/s | $\approx 3.0\times10^5$ | Strong PQC performance |
| | | | | SPHINCS+-128s | 247 ops/s | $\approx 1.4\times10^7$ | Not broadcast-friendly |
| [130] | HE OBU (WAVE stack) | NVIDIA Jetson AGX Xavier (2.2 GHz) | Sign + Verify | ECDSA-P256 | 3.9 ms / 300 msgs | $\approx 2.86\times10^4$ cycles/msg | End-to-end WAVE pipeline |

| Ref. | Platform Class | Device (Clock Speed) | Operation | Cryptographic Algorithm | Measured Performance (Latency/ Throughput) | Normalized Cycles / Verify (approx.) | Notes |
|---|---|---|---|---|---|---|---|
| [130] | HE OBU (WAVE stack) | NVIDIA Jetson AGX Xavier (2.2 GHz) | Sign + Verify | Falcon-512 | 12.7 ms / 300 msgs | ≈ $9.31\times10^4$ cycles/msg | End-to-end WAVE pipeline |

Notes- Cycles/verify: approximate normalized cycle count per verification, computed as (clock frequency ÷ measured operations per second) or (per-message latency × clock frequency); HE OBU: high-end OBU class; LE OBU: low-end OBU class; OBU: On-Board Unit; ops/s: verifications per second; Ref.: reference number; RSU: Roadside Unit.

### *5.3.3 Protocol and Regulatory Migration Challenges*

#### *5.3.3.1 Hybrid Cryptographic Migration Mechanisms and Trade-offs*

**Migration Rationale and Hybrid Deployment Models**: Transitioning V2X systems to PQC infrastructures creates operational challenges due to mixed legacy and next-generation devices (OBUs/RSUs), strict latency limits for safety-critical ITS applications, and backward-compatibility requirements with existing ECC-based V2X security protocols, certificate formats, and SCMS/PKI infrastructures. Hybrid cryptography addresses this by combining classical and PQC algorithms within messages or credentials so legacy receivers can validate classical signatures while PQC-capable receivers perform dual verification. For example, ECDSA-Falcon hybrid designs support backward compatibility and post-quantum validation on upgraded platforms [31,131,132]. Reported field and simulation studies show partially hybrid Falcon-based authentication adds only sub-millisecond overhead (≈0.25-0.6 ms per BSM) [131], remaining within the typical safety-message timing budget of 100 ms [17,131], and bandwidth impact can be reduced by attaching PQC artifacts mainly during certificate provisioning rather than with every BSM [126,131]. Hybrid V2X authentication models are typically grouped into three types: (i) true hybrid, where both classical and PQC signatures are attached and must be verified; (ii) backward-compatible hybrid, where dual signatures are sent but classical-only verification by legacy receivers is sufficient; and (iii) partially PQC hybrid, where messages use classical signatures while PQC is applied periodically to certificates or trust anchors [126,131]. These models balance bandwidth overhead, legacy compatibility, and long-term quantum security, with true hybrid designs generally preferred for safety-critical deployments once PQC verification support is widely available.

**Verification Policy and Signature Binding Requirements**: The security of hybrid deployments depends strongly on receiver-side verification policy and signature binding rules. Systems using any hybrid deployment model must define which signature components are mandatory and how validation outcomes are combined. If policies accept either classical or PQC signatures, downgrade attacks become possible in which PQC artifacts can be stripped, and classical-only validation is forced; in contrast, strongly bound (non-separable) hybrid signatures cryptographically link classical and PQC components so that removing one invalidates the entire signature object, thereby eliminating this downgrade path [133]. Experimental and protocol studies indicate that when hybrid signatures are created as a single combined object, an attacker would need to break both the classical and the PQC algorithms simultaneously to successfully forge a signature. Phased migration designs therefore protect the most critical trust anchors (such as CA signatures) with PQC first, while temporarily allowing limited classical signing during early deployment to maintain compatibility and forward security [131,134].

**Implementation and Parsing Complexity:** Hybrid cryptographic formats make software and message parsing complex because receivers must handle certificates and signatures that contain both classical and PQC keys and fields, and must run multiple verification paths. PQC keys and signatures are usually larger and variable in size, so implementations need flexible length handling and dynamic buffers, which increases the chance of parsing and memory-management errors in resource-constrained OBUs and RSUs. While certificate frameworks, such as X.509 and IEEE 1609.2, can support hybrid credentials, they increase certificate size and require more complex encoding and decoding logic, expanding the potential attack surface [132,135]. Large PQC signatures, such as SPHINCS+ and Dilithium, may exceed packet-size limits, which leads to message fragmentation and split-signature verification approaches that introduce additional parser state, buffering logic, ordering checks, and error-handling paths in OBU software stacks [130]. Consequently, hybrid V2X deployments typically require hardened parsers, strict bounds checking, multi-library cryptographic support, and rigorous memory management practices to avoid introducing new implementation vulnerabilities while integrating PQC.

**Measured Overhead and Capacity Impact:** Hybrid credentials and signatures increase bandwidth, storage, and processing load because both classical and PQC artifacts must be transmitted and verified. Classical V2X certificates are usually a few hundred bytes (≈330 bytes reported in experimental studies), whereas PQC or hybrid credentials are much larger, increasing packet size and channel usage, and reducing the number of authenticated messages that can be handled in dense traffic conditions [132,133]. Falcon-based hybrid authentication experiments show measurable capacity reduction, where partially hybrid certificate-protected settings reduce the number of supported authenticated vehicles per 100 ms interval (e.g., from about 165 to 101, approximately a 39% drop) under DSRC assumptions [130]. Hybrid migration also increases operational cost for SCMS and edge devices (OBUs/RSUs), because dual credential sets, multiple verification paths, and larger cryptographic libraries must be maintained, which can strain embedded processors and storage unless supported by hardware acceleration and optimized implementations [132,134]. Despite these overheads, hybrid schemes remain the most practical step-by-step approach for backward-compatible PQC transition in V2X systems.

##### *5.3.3.2 Pseudonym Certificate Management in PQC-Enabled SCMS*

Pseudonym certificate management becomes significantly more demanding under PQC because credential sizes and distribution costs increase. Classical V2X deployments rely on frequent pseudonym rotation to preserve privacy, but PQC public keys and signatures increase the size of certificate artifacts, increasing storage, caching, and distribution overhead. Under IEEE 1609.2-based ecosystems, vehicles may require about 20 pseudonym certificates per week, resulting in thousands per year [136]. When PQC or hybrid credentials are used, this scaling factor directly amplifies SCMS bandwidth and storage requirements.

SCMS infrastructures, therefore, need to accommodate higher certificate provisioning volume, larger artifact sizes, and longer verification times. Hybrid certificates that embed both classical and PQC keys and are signed by a common CA have been proposed to maintain backward compatibility while reducing implementation overhead. Reported designs indicate that such hybrid certificate strategies can reduce reissuance frequency while preserving interoperability, but they still increase per-certificate size and lifecycle management complexity [31,132]. As a result, PQC migration at the certificate-management layer primarily poses scalability and logistics challenges.

#### *5.3.3.3 Protocol-Level Upgrades for PQC Compatibility*

PQC key and signature sizes require protocol-level adaptations across vehicular communication stacks because larger public keys, signatures, and certificates directly increase message size and security payload overhead. Existing V2X security protocols, such as IEEE 1609.2, were originally optimized for ECC-based artifacts and therefore assume comparatively compact certificate and signature structures. When PQC or hybrid credentials are introduced, certificate-carrying security payloads can approach or exceed typical packet-size assumptions, requiring fragmentation-aware handling and revised encoding practices [130,131].

Certificate encoding format plays a measurable role in overhead. IEEE 1609.2 already supports implicit certificates, which reduce certificate-carrying SPDU size compared to explicit certificates in the classical setting (reported values are approximately 226 bytes with implicit certificates versus about 330 bytes with explicit certificates) [131]. However, comparable PQC implicit certificate constructions are not yet standardized or widely available [82], so current PQC and hybrid deployments primarily rely on explicit certificate formats, limiting the reduction in overhead when large PQC public keys are embedded [131].

Large PQC signatures directly stress the limits of packetization in secured V2X messaging because IEEE 1609.2 security envelopes add certificate and signature overhead on top of already-constrained link-layer packet sizes. Under ETSI C-ITS secure message profiles, the network MTU is commonly reported around 1,492 bytes, with roughly ≈1,428 bytes available for payload and security fields [30], while DSRC-based profiles are often cited with payload limits around ≈2,300 bytes [131]. Therefore, certificate-carrying IEEE 1609.2 or ETSI C-ITS SPDUs can quickly approach or exceed single-packet limits when PQC artifacts are used [131]. In this constrained envelope, compact PQC signatures such as Falcon are generally favored because they are more likely to fit within a single protected message, whereas larger signatures such as Dilithium frequently exceed secure payload limits and therefore require fragmentation and multi-frame transmission. Accordingly, hybrid and PQC V2X proposals describe fragment-aware payload formats, sequence identifiers, and segmented verification procedures to enable reliable reconstruction and validation across multiple WAVE Short Messages [130,131]. These size effects also propagate upward and downward in the protocol stack. For example, SAE J2735 message definitions (e.g., BSM structures), while not cryptographic themselves, may require variable-length payload markers to accommodate an estimated ≈35-40% payload growth from larger security artifacts while preserving backward compatibility with legacy DSRC stacks, and 3GPP sidelink mode-4 scheduling and resource allocation logic may require tuning to handle larger authentication headers under dense traffic conditions [131].

#### *5.3.3.4 Regulatory Ambiguity and Certification Overhead*

Regulatory uncertainty remains a significant barrier to the widespread adoption of PQC in vehicular and infrastructure systems. Current cybersecurity regulations, such as ISO/SAE 21434 and UN R155, define cryptographic safeguards in terms of state-of-the-art protections, yet fall short of specifying mandatory timelines, benchmarks, or migration requirements for PQC readiness [87,88]. As a result, manufacturers pursuing PQC-enhanced systems need to independently justify their implementations by submitting detailed migration plans, performance benchmarks, and compatibility documentation to meet certification requirements.

To compensate for these regulatory gaps, emerging practices, such as Quantum Risk Assessments (QRAs) and post-quantum migration readiness evaluations, are informed by guidance from NIST's security analyses (e.g., MAXDEPTH) [15]. In addition, the transition timelines outlined in the U.S. National Security Agency's (NSA) CNSA 2.0 framework [137,138] highlight the urgency of preparing for cryptanalytically relevant quantum computers (CRQCs).

In parallel with these risk-assessment efforts, certification submissions must include HSM Attestation Reports demonstrating FIPS 140-3 Level 3 compliance for programmable accelerators capable of securely executing PQC operations, including floating-point instructions required for Falcon verification [29,102,139]. To ensure interoperability, vendors also need to provide hybrid PKI validation kits that demonstrate secure cross-certification between classical (ECDSA P-256) and PQC trust anchors (Dilithium-2, Falcon-512) while keeping verification latency <2 ms [132]. Although NIST IR 8545 outlines algorithmic status and performance, it lacks domain-specific automotive deployment metrics [60]. Thus, real-world validation and certification of PQC remain key technical hurdles for large-scale V2X and ITS integration.

### *5.3.4 Advanced Post-Quantum Cryptographic Primitives for Future V2X Systems*

As vehicular systems become more connected and automated, post-quantum security needs extend beyond core cryptographic primitives such as signatures and KEMs to include privacy-preserving authentication, collaborative computation, and distributed trust mechanisms for ITS environments. Accordingly, this subsection reviews advanced PQC primitives and V2X-focused constructions, including group and anonymous credentials, lattice-based conditional privacy-preserving authentication, zero-knowledge proofs, homomorphic encryption, and secure multi-party computation, along with recent V2X-oriented protocol designs and evaluation frameworks that demonstrate their practical applicability.

#### *5.3.4.1 Group Signatures and Anonymous Credentials*

Group signatures and anonymous credential systems enable vehicles to authenticate attributes or sign messages without revealing their identities, thereby addressing the fundamental trade-off between accountability and privacy in ITS. Traditional schemes based on bilinear pairings, RSA, or ECC are vulnerable to quantum attacks, which motivates the development of PQC alternatives built on lattice and hash-based algorithms [140].

Group signatures allow any member of a defined group to sign on behalf of the group while preserving anonymity unless a designated authority opens the signature to identify the signer. Boneh et al. developed an Enhanced Privacy ID (EPID) group signature scheme that can support a group of over 1 million members, with reported signature sizes on the order of megabytes (≈3.45 MB) at 128-bit post-quantum security [141]. This scheme is primarily suited for attestation scenarios where application data dominates cryptographic overhead. Similarly, hash-based ring signatures, which are decentralized group-style signatures without a central authority, enable vehicles to prove message authenticity from valid participants without disclosing their identities, mitigating ubiquitous tracking risks [142].

Anonymous credential (AC) and lattice-based conditional privacy-preserving authentication (CPPA) schemes support privacy-preserving vehicular authentication by enabling attribute or message validation without revealing vehicle identity. Recent vehicular ad hoc network (VANET)-targeted PQC designs use lattice assumptions, dynamic pseudonyms, and batch verification to achieve scalable, anonymous authentication with revocation, as well as decentralized verification models suitable for large-scale V2X deployment [140,143–145]. Lightweight CPPA constructions emphasize reduced computation and message overhead for constrained vehicular nodes [144], while multi-aggregator architectures distribute verification across cooperating RSUs to improve scalability under dense traffic and report verification-delay reductions exceeding 50% relative to single-aggregator baselines [143]. Edge-assisted CPPA schemes further reduce redundant per-message verification through RSU-side batch validation and revocation support [145].

Furthermore, some approaches combine lattice-based authentication with aggregate signatures and decentralized trust, including blockchain-supported V2V authentication under module-lattice assumptions with aggregate verification to lower verifier workload while preserving conditional privacy and integrity [146,147]. Recent work also extends lattice-based privacy-preserving authentication to structured vehicular sub-networks such as autonomous platoons, where membership admission and join authentication become high-value attack targets. For example, a post-quantum platoon access authentication design by Xu et al. secures the initial platoon join and membership establishment phase using PQC-based authentication combined with signal-level protection measures, aiming to maintain reliable, low-latency communication under both quantum and cyber-physical interference attacks [148]. In addition, the Quantum-Safe Vehicular Platoon (QSVP) mutual authentication protocol applies a split-trust model in which a PQC KEM (Kyber) secures vehicle-facing links, while quantum key distribution-style keying is reserved for stationary RSU-server backbone links, improving scalability while preserving quantum-safe protection for infrastructure trust anchors [149].

*5.3.4.2 Zero-Knowledge Proofs for Privacy-Preserving Authentication*

A zero-knowledge proof (ZKP) enables vehicles to prove possession of valid credentials without revealing the underlying secret information [150]. ZKP-based authentication supports privacy-preserving vehicular applications, including anonymous toll collection, conditional access control, and location privacy. However, post-quantum ZKP systems face a fundamental trade-off. For example, lattice-based constructions offer quantum resistance but incur substantial proof sizes (120-724 KB) [151]. Recent implementations demonstrate that lattice-based ZKPs can be practical for vehicular systems under limited or non-real-time constraints. For example, a post-quantum anonymous credential scheme using ML-DSA (Dilithium) with the Poseidon hash function [152] (replacing SHA-3 to reduce multiplicative circuit depth) achieves proof sizes of 75-85 KB and generation times of 0.3-5 seconds [151]. While these latencies exceed real-time V2X requirements, they are more suitable for periodic authentication tasks, such as weekly pseudonym certificate provisioning [127]. For V2X-specific applications, Chaudhry et al. propose ZKPs for vehicular health data sharing in V2Healthcare (V2HX) systems, where vehicles authenticate to healthcare entities without revealing driver identities using optimized lattice-based constructions [153]. However, a critical challenge in ZKP systems is efficiently constructing zero-knowledge proofs of the correct evaluation of conventional hash functions (e.g., SHA-3) within ZKP circuits, since their circuit representations impose many constraints and thereby introduce substantial computational and communication overhead [151].

#### *5.3.4.3 Privacy-Preserving Collaborative Computation, Secure Analytics, and Encrypted Aggregation*

Secure multi-party computation (MPC) protocols enable multiple vehicles or infrastructure entities to jointly compute functions over their private inputs without revealing those inputs to each other or to a central party [154]. Representative applications include collaborative route planning, distributed traffic optimization, and privacy-preserving federated learning for autonomous driving models. Secret sharing, a foundational MPC primitive, has received particular attention for vehicular federated learning, where vehicles collaboratively train machine learning models without sharing raw data. In such settings, model updates or parameters are divided into cryptographic shares and processed collectively, rather than exchanged in plaintext. Post-quantum secret sharing schemes based on lattice problems can enable vehicles to split model parameters into shares, where each share individually reveals no information and only a qualified subset can reconstruct the original value, thereby preventing privacy leakage even against quantum adversaries [155].

Homomorphic encryption (HE) provides a complementary approach for privacy-preserving federated learning by allowing aggregation and limited computation to be performed directly on encrypted model updates rather than on secret shares [156]. In HE-based federated learning, vehicles encrypt local gradients or model parameters before transmission, and the aggregator computes the global update over ciphertexts without accessing plaintext data. This approach has been demonstrated in vehicular and Internet-of-Vehicles (IoV) settings, including HE-enabled federated learning frameworks for privacy-preserving intrusion detection, where encrypted model updates from vehicles are securely aggregated at the server side [157]. HE-based aggregation reduces information leakage from intermediate model updates and serves as a complementary approach to MPC and secret-sharing techniques in quantum-safe collaborative learning architectures.

Recent vehicular and edge-centric PQC frameworks demonstrate end-to-end secure analytics pipelines in addition to standalone cryptographic primitives. For example, an edge-IoT PQC security framework reports a hybrid post-quantum deployment using Kyber and Dilithium with AES-GCM across a device-edge-cloud workflow, reporting measured performance gains of about 18.6% higher throughput, 22.4% lower latency, and 41.7% lower energy consumption, with per-transaction cryptographic overhead below 7.3 ms under the evaluated edge configuration [158]. A dedicated PQ-V2X dataset has been introduced, containing quantum-safe vehicular communication traces generated using multiple NIST PQC algorithms (including Kyber, Dilithium, Falcon, and SPHINCS+), enabling reproducible benchmarking of intrusion detection systems (IDSs) and analytics models on PQ-secured traffic rather than classical-only traces [159]. Complementarily, a machine learning (ML)-driven vehicular cloud IDS framework combined with federated quantum key cryptography demonstrates that quantum-safe key establishment can be integrated with ML-based IDS pipelines to support authenticated, privacy-preserving V2V/V2X communication under encrypted cloud-assisted settings [160].

#### *5.3.4.4 AI-Assisted Adaptive and Context-Aware Cryptographic and Protocol Framework*

Beyond fixed-algorithm deployments, recent studies propose AI-assisted adaptive post-quantum cryptographic control frameworks for highly dynamic vehicular environments, where network conditions, mobility patterns, and device capabilities vary continuously [161,162]. These frameworks introduce cryptographic agility by enabling runtime selection of PQC schemes and parameters rather than relying on static algorithm configurations. For example, a context-aware adaptive PQC (CAAP) framework for 6G V2X environments dynamically selects among schemes, such as Kyber, Dilithium, SPHINCS+, and McEliece, based on mobility patterns, channel quality, and computational constraints, using multi-objective optimization and reinforcement learning to guide runtime decisions [162]. Trace-driven vehicular channel and mobility evaluations using CAAP report up to 27% reduction in end-to-end cryptographic latency while preserving security during algorithm switching. Similar context-aware PQC selection strategies could also be applied under adverse radio conditions, such as bad weather or severe channel degradation, where link quality and latency constraints change rapidly.

## 5.4 Research Gaps in PQC Integration for ITS Communication Interfaces

The preceding subsections (5.1-5.3) examined PQC integration across in-vehicle, OEM-specific, and V2X communication links, highlighting persistent challenges involving protocol fragmentation, edge-device performance, and regulatory uncertainty. As summarized in **Fig. 4**, these challenges correspond to seven key research gaps (RG4-RG10) spanning in-vehicle, OEM-specific, and V2X communication links within the ITS ecosystem. Although mitigation strategies, such as hybrid cryptography, certificate compression, and fragmentation-aware transport, offer a partial solution, several critical research gaps continue to hinder scalable, standards-compliant PQC deployment within ITS ecosystems.

- **RG4** addresses the lack of protocol-agnostic fragmentation mechanisms capable of efficiently supporting large PQC payloads, such as SPHINCS+ (SLH-DSA) signatures, which reach 49 KB. In in-vehicle networks, legacy protocols like CAN-FD, constrained by a 64-byte MTU, are not equipped to transmit large PQC artifacts without custom transport-layer enhancements. Meanwhile, in V2X environments, even though higher MTUs are supported, the fragmented transmission of large PQC payloads introduces delays, increases the risk of packet loss, and may result in failed reassembly under congested traffic conditions or DoS attacks. Both contexts highlight the urgent need for standardized fragmentation support across heterogeneous vehicular protocols.
- **RG5** highlights the lack of energy-optimized PQC implementations for sub-100 MHz automotive ECUs, where schemes, such as Kyber and Dilithium, currently encounter unsustainable power and memory overheads.
- **RG6** reflects the need for adaptive hybrid authentication frameworks that ensure interoperability between classical and PQC-based systems. Without such mechanisms, legacy vehicles may reject PQC-augmented messages, causing cross-certification failures or message misinterpretation during transitional deployments.
- **RG7** points to the lack of standardized compression techniques for large PQC artifacts, such as SPHINCS+ signatures, which could mitigate fragmentation-induced latency and jitter in bandwidth-constrained networks. Current ITS standards do not define compression or adaptation methods for PQC payloads.

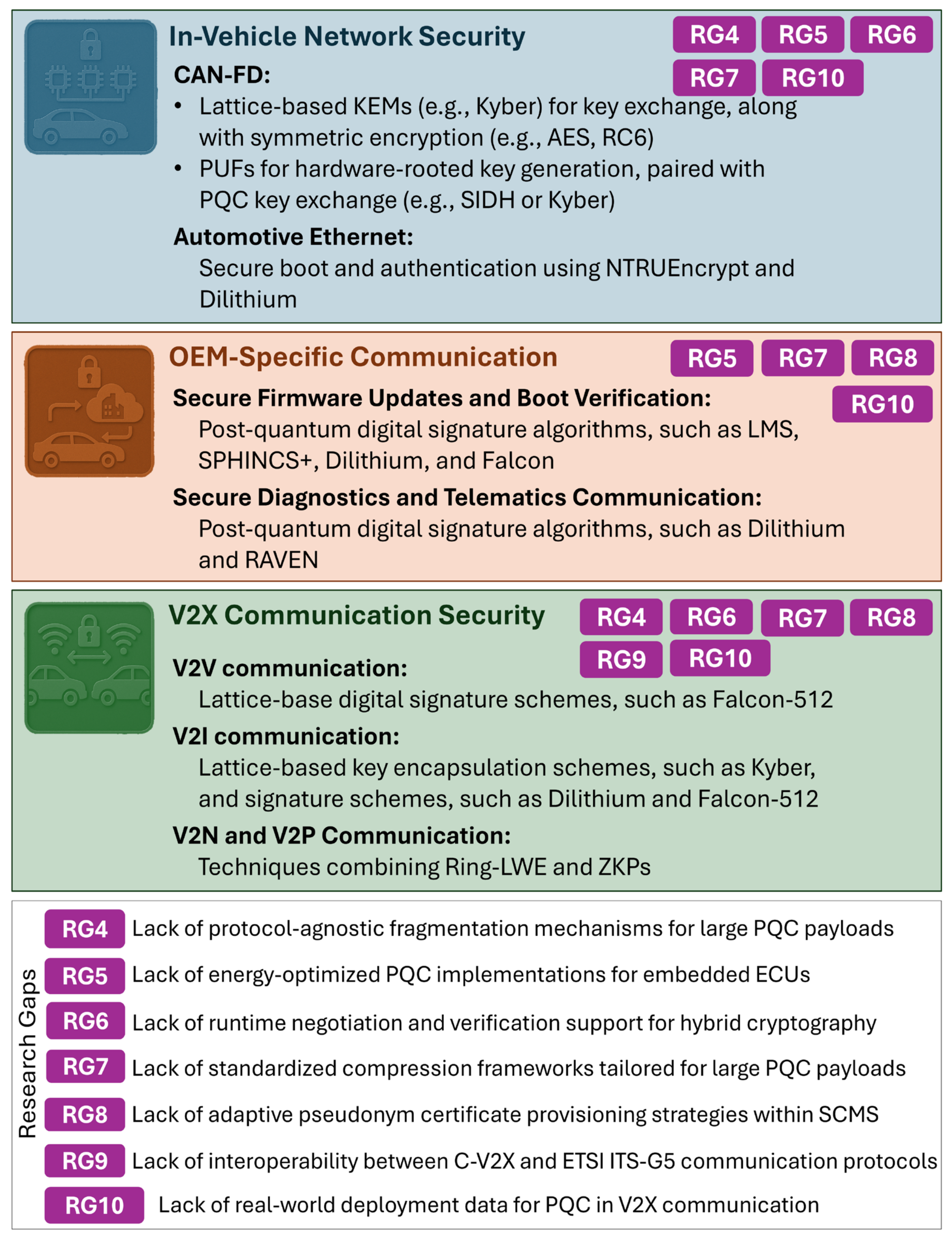

**Fig. 4.** Mapping of research gaps (RG4-RG10) across in-vehicle, OEM-specific, and V2X communication domains. Abbreviations- KEM: Key Encapsulation Mechanism; LWE: Learning with Errors; OEM: Original Equipment Manufacturer; PQC: Post-Quantum Cryptography; PUF: Physically Unclonable Functions; V2I: Vehicle-to-Infrastructure; V2N: Vehicle-to-Network; V2P: Vehicle-to-Pedestrian; V2V: Vehicle-to-Vehicle; V2X: Vehicle-to-Everything; ZKP: Zero-Knowledge Proof.

- **RG8** captures the lack of scalable pseudonym certificate provisioning strategies within SCMS. Larger PQC credentials increase memory requirements, reduce rotation frequency, and complicate revocation and real-time distribution.

- **RG9** denotes the absence of harmonized protocol compatibility for PQC-enhanced messages across heterogeneous V2X stacks, such as C-V2X and ETSI ITS-G5, which may hinder global interoperability and deployment in mixed-mode environments.
- **RG10** reflects the absence of real-world deployment data for PQC in automotive settings. Minimal empirical evidence exists on how PQC algorithms perform under operational automotive conditions, such as temperature variation, vibration, electromagnetic noise, and constrained processing. This lack of deployment feedback forces system designers to rely solely on simulation or lab evaluations, increasing the risk of underperformance or vulnerabilities in real-world scenarios.

### 5.5 Concluding Remarks of Section 5

To summarize the in-vehicle subsection (Section 5.1), PQC feasibility depends strongly on protocol bandwidth, timing constraints, and ECU hardware capability. Low-payload and schedule-constrained buses, such as CAN and FlexRay, typically require hybrid approaches and higher-layer fragmentation, whereas Automotive Ethernet offers a more practical near-term path for broader PQC integration. Reported benchmarks indicate that lattice-based schemes are generally more deployable than hash-based or code-based alternatives under embedded resource limits.

To summarize the OEM and cloud communication subsection (Section 5.2), PQC integration is comparatively more practical because these channels operate over higher-bandwidth links and updateable software stacks. PQC can be incorporated into secure boot, OTA update pipelines, and backend authentication using hybrid migration models with fewer latency constraints than in broadcast vehicular channels. The literature indicates that backend infrastructures can serve as early PQC deployment anchors that support broader vehicular trust chains. However, this shift increases certificate lifecycle and key management complexity at scale.

To summarize the V2X communication subsection (Section 5.3), PQC integration is mainly constrained by signature size growth, certificate overhead, and strict safety-message latency requirements. Larger PQC artifacts increase fragmentation risk and verification load, making hybrid signature and dual-certificate strategies the most practical transition approach. Across studies, authentication and certificate handling overhead emerges as a greater bottleneck than encryption cost. PQC adoption in V2X, therefore, depends on coordinated protocol and certificate architecture adaptation, together with message-format and transport-layer adjustments. In addition, advanced PQC primitives, including anonymous credentials, group signatures, ZKPs, homomorphic encryption, and MPC-based collaborative analytics, expand the security and privacy capabilities of V2X systems, but introduce additional proof-size, computation, and protocol-complexity considerations that must be accounted for in future standards and deployment profiles.

This subsection-level synthesis is based on heterogeneous benchmarks, prototype implementations, and simulation-driven studies reported across different hardware platforms, so performance comparisons should be interpreted as directional rather than directly comparable. Large-scale field deployments and production-grade PQC vehicular stacks remain limited, and vendor-specific accelerator and firmware results are often not publicly available. In addition, this review focuses on communication-layer feasibility and does not include detailed cost, supply-chain, or lifecycle maintenance analysis for automotive PQC deployment. The next section, therefore, examines implementation-level and physical attack considerations that affect the practical security of PQC-enabled vehicular systems.

# 6. PHYSICAL LAYER VULNERABILITIES: SIDE-CHANNEL AND FAULT INJECTION ATTACKS ON PQC

Although PQC algorithms resist quantum attacks, they remain vulnerable to physical-layer threats, primarily side-channel attacks (SCA) and fault injection (FIA) attacks. These attacks exploit hardware leakage and induced faults, respectively, rather than cryptographic weakness and are especially critical in embedded automotive environments, where PQC algorithms need to operate on resource-constrained, real-time systems. This section reviews documented SCA and FIA vulnerabilities across PQC schemes, evaluates existing countermeasures in automotive-grade hardware, and outlines several corresponding research gaps for secure PQC deployment in ITS applications, with emphasis on how physical-layer compromise of PQC modules can propagate upward into safety-critical ITS functions, such as secure boot, OTA updates, and authenticated V2X messaging.

## 6.1 Side-Channel Attacks in Embedded and Automotive Contexts

Lattice-based PQC schemes, including CRYSTALS-Kyber and CRYSTALS-Dilithium, have demonstrated susceptibility to power, timing, and electromagnetic (EM) side-channel leakage during operations such as polynomial multiplication and the Number Theoretic Transform (NTT) [163,164]. Correlation and differential power analyses (CPA/DPA) have successfully recovered Kyber secret keys by exploiting NTT computations, even under masking [163,165–167]. Deep learning-assisted CPA has further enabled full key recovery during decapsulation on embedded platforms [166,167]. Dilithium, though avoiding floating-point arithmetic, remains vulnerable to DPA during signature generation [163,165,168]. Convolutional neural network (CNN)-based profiling attacks can extract secrets from masked implementations of Dilithium without physical access, posing a risk to unshielded ECUs [168]. Falcon, reliant on floating-point Fast Fourier Transforms (FFTs), exhibits leakage in its Gaussian sampler and FFT pipeline, which can be exploited via cache and power analyses [169,170]. Similarly, HQC decoders leak through timing and power channels, where Oracle Template-Plaintext Checking Attacks (OT-PCAs) achieve private key recovery with a few oracle queries, validated on Cortex-M4 platforms [171,172]. From an ITS perspective, such leakages can enable the extraction of long-term device keys used for V2X authentication or secure firmware validation, allowing attackers to inject forged messages or malicious updates that directly affect vehicle and infrastructure safety behavior.

## 6.2 Fault Injection Attacks on PQC

Fault injection attacks (FIAs), including clock glitches, instruction skips, and rowhammer-induced bit flips, can alter control flows or corrupt cryptographic outputs. For example, PQ-Hammer has demonstrated full key recovery on Dilithium and Falcon by exploiting DRAM bit flips, posing serious risks for automotive ECUs with limited electromagnetic shielding [173]. Kyber is similarly vulnerable during decapsulation, where loop-abort or instruction-skip faults in NTT and decoder logic enable secret key extraction [163,174,175]. Dilithium's deterministic signature generation can also be compromised by single-instruction skips that cause nonce (a one-time random value used to ensure each signature is unique) reuse and key recovery [163]. In ITS, these fault-induced key recoveries translate into practical risks, such as bypassing ECU trust anchors, disabling secure boot checks, or forging authenticated safety messages, which makes FIA resilience a system-level safety concern that extends beyond hardware security.

### 6.3 AI-Assisted SCAs and Enhanced Profiling

The adoption of machine learning (ML) and deep learning (DL) has significantly enhanced the capabilities of side-channel attacks [167,168,176–179]. CNN-based and Long Short-Term Memory (LSTM)-based profiling now enables key recovery with fewer than 500 traces, compared to thousands previously required [178]. A recent study demonstrated blind CNN-based key recovery on Kyber without knowledge of inputs or fault injection, suggesting that such attacks may be applicable beyond controlled laboratory settings [167]. Similar CNN architectures trained on leaked traces from embedded systems have successfully broken Dilithium [179,180]. A 2025 doctoral study further demonstrated hybrid DL models that reduce trace requirements and enhance reliability under low-noise conditions [176]. These AI-assisted approaches significantly lower the effort and data needed for successful attacks, increasing the risk of side-channel exploitation in ITS environments.

### 6.4 Countermeasure Challenges in Embedded Platforms

For transportation engineers, the limitations of currently proposed side-channel and fault-injection countermeasures directly affect whether PQC protections assumed in system safety and security designs actually hold under field attack conditions. Although countermeasures, such as masking, shuffling, and constant-time logic, have been proposed, many fail under modern ML-assisted profiling attacks [165–167,181]. For example, CNN-based trace alignment breaks Kyber's coefficient shuffling [167], while Dilithium's masking leaks indices due to its interaction with rejection sampling [165,168]. In automotive ECUs, stringent real-time, power, and isolation constraints further limit the effectiveness of protection. Recent hardware-friendly solutions, such as probabilistic masking, adaptive shuffling, and multi-core isolation, improve resilience but remain impractical for legacy or resource-constrained platforms [166,167,181].

### 6.5 Real-World Implications for Automotive PQC Deployment

As PQC becomes integral to secure boot, OTA updates, and V2X communication, resilience against side-channel and fault-injection vulnerabilities is critical for both cybersecurity and functional safety assurance in deployed ITS platforms. Validated attacks on ARM Cortex-M4, Cortex-A55, and RISC-V cores confirm the practicality of such threats in automotive contexts [165,175,177,178]. Vehicular systems face heightened exposure due to physical access to telematics units, long service lifespans, and shared memory buses, while limited shielding and strict real-time constraints restrict countermeasure adoption. Without robust physical-layer protections, PQC integration could introduce new vulnerabilities. Future work should establish automotive-specific side-channel benchmarks, integrate lightweight countermeasures into ISO/SAE 26262 (an international standard for the functional safety of electrical and electronic systems in road vehicles [182]) and AUTOSAR, and develop automated frameworks for evaluating PQC fault and side-channel resilience. **Table 13** summarizes key physical-layer attack methods, affected algorithms, and implications for automotive systems.

**Table 13**
**Summary of physical-layer attacks on PQC schemes**

| References | Scheme | Attack Type | Attack Technique | Impact |
|---|---|---|---|---|
| [137–139,163–167,174,175,177,181] | Kyber (ML-KEM) | Side-channel | Power/EM leakage in NTT; CPA/DPA; CNN/LSTM profiling | Secret/session key recovery |
| [163,175] | Kyber (ML-KEM) | Fault injection | Instruction-skip/loop-abort during decapsulation/decoder | Secret key recovery |
| [163,168,179,180] | Dilithium (ML-DSA) | Side-channel | DPA; CNN profiling during signing (rejection sampling/index leaks) | Nonce leakage to forgery |
| [169] | Falcon | Side-channel | Leakage in Gaussian sampler and FFT (cache/power) | Key extraction/forgery |
| [171,172] | HQC | Side-channel | Decoder leakage; OT-PCA oracle; timing/power analysis | Private key recovery |
| [173] | PQC (general) | Fault injection | Rowhammer (PQ-Hammer) bit-flip induced faults | Full key recovery (case-dependent) |

Notes- CNN: Convolutional Neural Network; CPA: Correlation Power Analysis; DPA: Differential Power Analysis; DSA: Digital Signature Algorithm; EM: Electromagnetic; FFT: Fast Fourier Transform; KEM: Key Encapsulation Mechanism; LSTM: Long Short-Term Memory; NTT: Number Theoretic Transform; OT-PCA: Oracle Template–Plaintext Checking Attack; PQC: Post-Quantum Cryptography.

## 6.6 Research Gaps in Physical-Layer Resilience of PQC for Automotive Systems

Despite progress in identifying side-channel and fault vulnerabilities, achieving robust, real-time physical-layer resilience in automotive environments remains challenging, and unresolved weaknesses at this layer can undermine higher-level ITS security and safety guarantees. As illustrated in **Fig. 5**, vulnerabilities and research gaps (RG11-RG13) span side-channel leakage, fault injection, AI-assisted profiling, and countermeasure limitations, linking specific PQC weaknesses, such as Dilithium's masking flaws, Kyber's trace misalignment, and Falcon's power and cache leakage, to their corresponding open challenges.

- **RG11** reflects the lack of standardized fault isolation and cryptographic redundancy mechanisms in hybrid PQC-classical systems. During transition phases, attackers may exploit weaker classical algorithms, such as ECDSA, while PQC schemes like Dilithium are concurrently deployed, leading to downgrade or forgery risks.
- **RG12** reflects the absence of cost-effective defenses against AI-assisted SCAs, including CNN-based and LSTM-based profiling. Traditional countermeasures, such as masking and coefficient shuffling, often fail under high-noise and resource-constrained automotive conditions.
- **RG13** highlights the lack of standardized fault-injection resilience requirements for PQC implementations in automotive systems that are certified for functional safety. Current frameworks, including ISO 26262 and AUTOSAR, do not address emerging attack vectors, such as rowhammer or instruction-skip faults, despite their demonstrated feasibility on automotive-grade ECUs [183].

## 6.7 Concluding Remarks of Section 6

PQC adoption does not eliminate implementation-layer risk and can introduce new side-channel, fault-injection, and profiling attack surfaces due to algorithmic structure and computational complexity. Reported studies demonstrate measurable leakage in lattice-based and hash-based PQC implementations under timing, power, electromagnetic, and AI-assisted profiling attacks when countermeasures are absent. The synthesis indicates that embedded automotive platforms need to treat PQC migration and physical-attack resistance as a joint design problem.

However, this review relies on published attack and countermeasure studies and does not include vendor evaluations or certified hardware module test results. Empirical attack success rates in real vehicular environments are rarely reported in the literature. Hence, the risk discussion here reflects demonstrated attack capability rather than how often such attacks occur in practice.

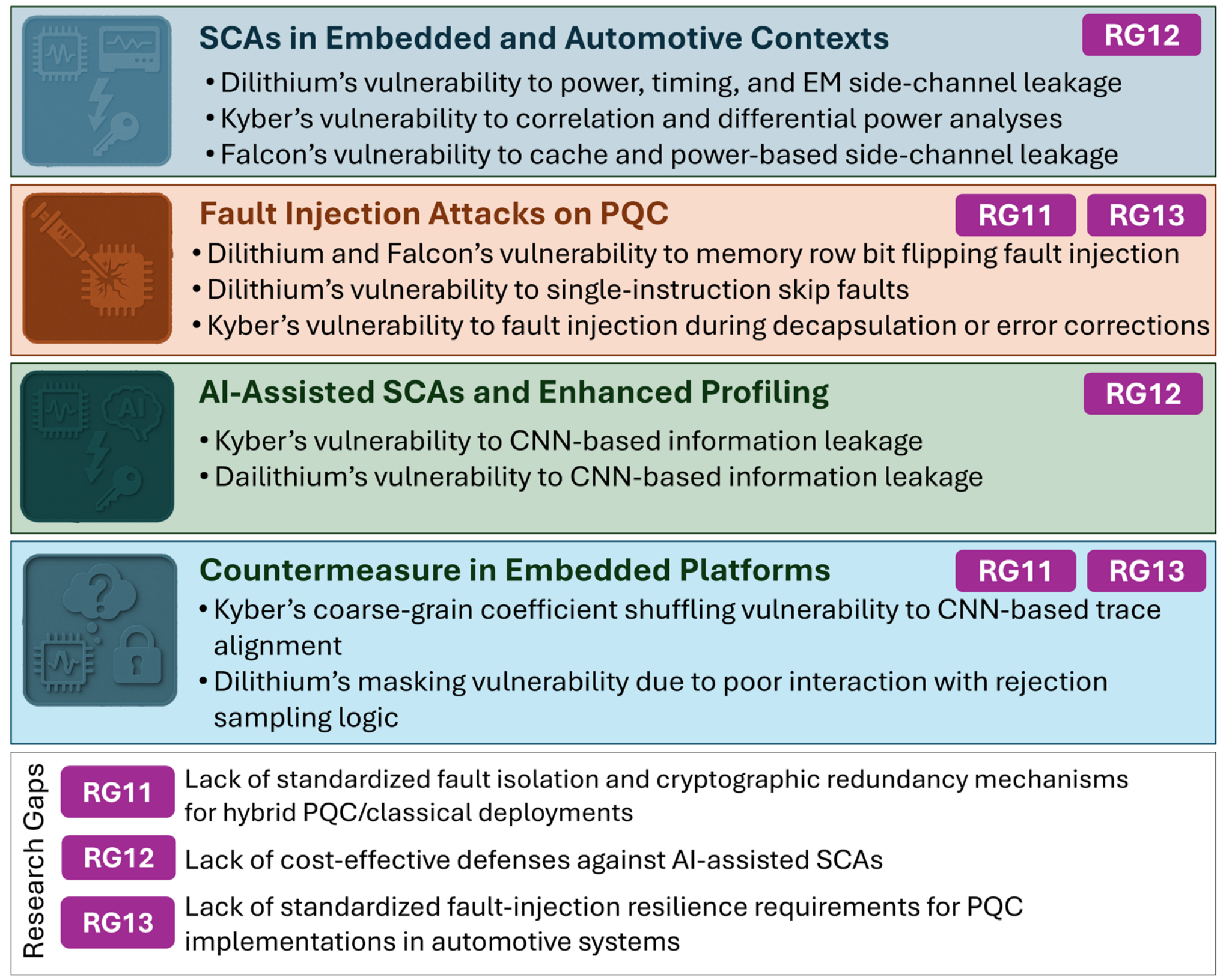

**Fig. 5.** Mapping of physical-layer vulnerabilities and research gaps (RG11-RG13) in PQC implementations for automotive systems. Abbreviations- CNN: Convolutional Neural Network; EM: Electromagnetic; PQC: Post-Quantum Cryptography; SCA: Side-Channel Attack.

## 7. FUTURE DIRECTIONS AND ROADMAP FOR PQC INTEGRATION IN ITS

This section consolidates the thirteen research gaps (RG1-RG13) identified earlier and outlines corresponding key future directions for scalable, interoperable, and quantum-resilient cryptography across the ITS ecosystem (**Table 14**). It is worth noting that not all identified research gaps have equal urgency or deployment impact. Some gaps, particularly certificate size expansion, protocol fragmentation limits, and hybrid interoperability (RG1-RG4, RG6), act as near-term deployment blockers because they directly affect message feasibility, standards compliance, and on-channel transmission limits in current ITS stacks. Others, such as embedded optimization and compression (RG5, RG7-RG9), pose scalability and efficiency challenges that can be improved incrementally across platform generations. A third group, including AI-assisted side-channel and fault-injection resilience (RG11-RG13), represents medium- to long-term hardening needs driven primarily by the evolution of attacker capabilities rather than by immediate deployment failure risk.

**Table 14**
**Research gaps and corresponding future research directions for PQC integration in ITS**

| Research Challenge Areas | Research Gap | Future Research Direction |
|---|---|---|
| Standard and Certificate Evolution | RG1: Lack of quantum-safe certificate formats in IEEE/ETSI standards | Extend existing protocols, such as IEEE 1609.2 and ETSI TS 103 097, to include PQC algorithms and support ASN.1 updates for multi-kilobyte keys |
| | RG2: Regulatory ambiguity on PQC adoption timelines and compliance benchmarks | Amend automotive cybersecurity and regulatory standards, such as ISO/SAE 21434 and UNECE WP.29, to define PQC adoption timelines and cryptographic compliance benchmarks |
| | RG3: Absence of cross-certification between classical and PQC trust anchors | Develop hybrid certificate frameworks enabling cross-certification between classical and PQC root authorities |
| | RG4: Lack of protocol-agnostic fragmentation mechanisms for large PQC payloads | Design fragmentation-resilient transport protocols and update specifications, such as SAE J2735 and 3GPP, for PQC compatibility |
| Embedded Optimization and Hybrid Cryptography | RG5: Lack of energy-optimized PQC implementations for embedded ECUs | Optimize PQC algorithms, such as Kyber and Dilithium, for ECUs using RISC-based processors or FPGA-based acceleration |
| | RG6: Lack of runtime negotiation and verification support for hybrid cryptography | Implement hybrid cryptography with dual-signature and partially hybrid certificate rotation schemes |
| Protocol Compression, Certificate Management, and Interoperability | RG7: Lack of standardized compression frameworks tailored for large PQC payloads | Develop standardized compression methods for PQC artifacts compatible with vehicular communication protocols, such as C-V2X and ETSI ITS-G5 |
| | RG8: Lack of adaptive pseudonym certificate provisioning in PQC-enabled SCMS | Adapt SCMS to support dynamic pseudonym provisioning, larger cryptographic credentials, and hybrid trust models |
| | RG9: Cross-stack interoperability challenges between C-V2X and ETSI ITS-G5 protocols | Design negotiation protocols and format-agnostic certificate containers for cross-stack interoperability |
| Real-World Deployment and Validation | RG10: Lack of real-world deployment data for PQC performance in V2X communication | Conduct pilot deployments with PQC-enabled OBUs/RSUs under real-world V2X conditions |
| Transition Security and Physical Attack Resilience | RG11: Lack of standardized fault isolation and cryptographic redundancy mechanisms in hybrid PQC/classical deployments | Develop and integrate trust precedence logic and safe-mode fallback mechanisms for hybrid deployments |
| | RG12: Lack of cost-effective defenses against AI-assisted SCAs | Harden PQC implementations with memory obfuscation, trace whitening, and control-flow randomization |
| | RG13: Lack of standardized fault-injection resilience requirements for PQC implementations in automotive systems | Add PQC-specific FIA test protocols to functional safety and automotive software frameworks, such as ISO 26262 and AUTOSAR, including instruction-level fault detection |

Notes- ECU: Electronic Control Unit; FIA: Fault Injection Attack; FPGA: Field-Programmable Gate Array; OBU: On-Board Unit; PKI: Public Key Infrastructure; RISC: Reduced Instruction Set Computer; RSU: Road-Side Unit; SCMS: Security Credential Management System; SCA: Side-Channel Attack; V2X: Vehicle-to-Everything.

## 7.1 Future Directions

### *7.1.1 RG1-RG4: Standard and Certificate Evolution*

To address RG1-RG3, vehicular communication and security standards, such as IEEE 1609.2, ETSI TS 103 097, and ISO/SAE 21434, must be updated to incorporate NIST-approved PQC algorithms and enable hybrid certificate infrastructures [31,87,88,131]. ASN.1 and certificate formats should be revised to support multi-kilobyte PQC keys, while UN R155/R156 regulations should formally recognize PQC-based root CAs. Furthermore, cross-certification mechanisms need to allow dual-signed certificates (ECC + PQC) to ensure continuity during migration [31].

To address RG4, distinct strategies are needed for in-vehicle and V2X communication domains. In in-vehicle networks, protocols such as CAN-FD, with an MTU of 64 bytes, are unable to carry large PQC artifacts, such as SPHINCS+ signatures, which can reach up to 49 KB. This calls for transport-layer segmentation and efficient reassembly mechanisms that meet ECU timing constraints without inducing latency or packet loss. In V2X systems, standards such as SAE J2735 and 3GPP NR-V2X must expand message size thresholds and integrate PQC references into upper-layer security profiles while maintaining the 100 ms latency budget for BSMs. Fragmentation-resilient transport protocols are also required to ensure reliable packet reconstruction under high-traffic conditions across DSRC, C-V2X, and hybrid stacks [131].

#### *7.1.2 RG5-RG6: Embedded Optimization and Hybrid Cryptography*

Addressing RG5 requires PQC implementations optimized for automotive hardware. Current ECUs, such as ARM Cortex-M processors, lack sufficient computational throughput for real-time PQC operations. Energy-efficient variants (e.g., Kyber512-90s), FPGA-based accelerators, or reduced-instruction-set designs should be explored to meet power and thermal limits in vehicles [30].

To resolve RG6, hybrid authentication schemes are required to enable seamless coexistence between classical and PQC algorithms. Dual-signature approaches that use both ECDSA and Dilithium provide backward compatibility during the transition, while intermittent transmission of PQC certificates reduces bandwidth overhead [31]. Agencies such as ANSSI (French National Cybersecurity Agency) and BSI (German Federal Office for Information Security) recommend similar hybrid migration paths [30]. In practice, classical signatures may secure routine BSMs, whereas PQC credentials are used during certificate updates [131]. Over time, these hybrid frameworks should evolve toward fully PQC-based cryptography once standards and validation systems mature.

Beyond core KEM and signature primitives, future embedded optimization efforts should also consider lightweight post-quantum privacy-preserving authentication mechanisms, such as lattice-based CPPA, group or anonymous credentials, and batch-verification schemes, which reduce verifier workload and communication overhead in dense vehicular environments [140,143–145]. These techniques can complement hybrid cryptography by improving scalability while preserving privacy and post-quantum assurance.

#### *7.1.3 RG7-RG9: Protocol Compression, Certificate Lifecycle Management, and Cross-Stack Interoperability*

To address RG7, PQC compression techniques can be standardized to reduce signature and key sizes for bandwidth-constrained networks, such as DSRC and C-V2X. Approaches, such as lattice coefficient pruning and polynomial sparsity encoding, can lower transmission overhead but require integration into IEEE 1609.2 and ETSI TS 103 097 to ensure consistent decoding and compatibility with pseudonym rotation [130,131]. In addition, compression mechanisms need to be introduced to preserve signature validity and certificate integrity while maintaining compatibility with pseudonym rotation and message signing requirements [131].

For RG8, the SCMS should evolve toward adaptive provisioning to handle larger PQC certificates with variable lifetimes and hybrid fields. Real-time feedback from vehicle density or communication load can optimize issuance and reduce storage strain. New lifecycle strategies, such as longer certificate validity, reduced rotation frequency, and cross-certification between classical and PQC CAs, will be required [30,31,131]. Pilot deployments, such as the hybrid SCMS prototype by Chen and Lin [132], already demonstrate the feasibility of efficiently combining ECC and PQC pseudonym certificates.

Addressing RG9 demands interoperability across heterogeneous V2X stacks (DSRC, C-V2X, ETSI ITS-G5). Hybrid deployments often face mismatches in message size, parsing inconsistencies, and limited algorithmic agility [31,131]. Future standards should introduce algorithm-negotiation extensions, format-agnostic certificate containers, and lightweight fallback mechanisms to ensure seamless coexistence of classical and post-quantum credentials across all ITS communication links [30,31,126,127,131,132]. In addition, emerging post-quantum anonymous credentials and ZKP-based authentication mechanisms may support selective disclosure and privacy-preserving certificate validation, especially for provisioning, tolling, and controlled-access services, although their current proof sizes and generation latencies limit use to non-real-time or periodic workflows [151,153].

#### *7.1.4 RG10: Real-World Deployment of PQC in V2X Systems*

Addressing RG10 requires real-world deployment of PQC in V2X systems beyond simulation, which often overlook practical constraints such as hardware contention, variable SNR, and network congestion [131]. Field trials using PQC-enabled OBUs and RSUs across varied environments (urban intersections, rural corridors, tunnels, highway mergers, etc.) will reveal implementation bottlenecks under realistic weather and mobility conditions. Mixed-fleet testing of classical cryptography and PQC-enabled communications can evaluate coexistence performance and establish key benchmarks, such as handshake latency, reliability, and packet fragmentation overhead [30,131]. Testing on automotive-grade ECUs (e.g., ARM Cortex-R, A-series) is critical to ensure production-level feasibility [30]. Insights from these deployments will guide PQC parameter tuning, cryptographic suite selection, and protocol adaptation for standards bodies like NIST, IEEE, ETSI, and ISO/SAE.

Future field trials should also extend beyond signature and KEM benchmarking to include advanced post-quantum secure analytics and collaborative mechanisms, such as homomorphic encryption, secure multi-party computation, and PQC-secured federated learning, as well as PQC-native datasets and IDS evaluation pipelines, to measure end-to-end privacy, latency, and resilience trade-offs under realistic workloads [157–160]. Context-aware adaptive PQC selection frameworks that dynamically switch algorithms based on channel and compute conditions should also be validated in live V2X environments [162].

#### *7.1.5 RG11-RG13: Transition Security and Physical Attack Resilience*

Addressing RG11 requires standardized mechanisms for fault isolation and redundancy during hybrid PQC/classical deployment. Without these, attackers may exploit weaker classical algorithms. Future systems should prioritize PQC signatures (e.g., Dilithium) over legacy ones (e.g., ECDSA) and include dual verification with fallback modes for safety-critical operations, such as braking or platooning [31].

For RG12, AI-driven side-channel attacks using CNN and LSTM profilers demand stronger protections, such as randomized memory access, control-flow obfuscation, and trace whitening. These defenses must be tested on automotive-grade platforms under realistic noise conditions [165,168,176–179,181].

RG13 highlights the absence of fault-injection resilience standards in frameworks like ISO 26262 and AUTOSAR [163,172,173,175]. PQC modules must be tested against electromagnetic, laser, and voltage faults using techniques such as instruction duplication and modular redundancy to ensure operational safety [177,181].

## 7.2 Roadmap for PQC Integration in ITS

Building on the strategies outlined in Sections 7.1.1 to 7.1.5, **Table 15** presents a phased roadmap for PQC adoption across the ITS ecosystem. The thirteen research gaps (RG1-RG13) are organized into short-, medium-, and long-term priorities defined relative to PQC standardization progress, vehicular standards revision cycles, automotive platform refresh patterns, and the quantum-hardware roadmaps discussed in Section 3.3.

The short term (within approximately 5 years) focuses on standards, certificate, and protocol-layer changes. These tasks mainly involve updates to standards documents, certificate profiles, and protocol specifications, and therefore require coordination across standards bodies, regulators, OEMs, and infrastructure operators. While such coordination and approval processes can be time-consuming, the required technical changes can be deployed on existing ECU, OBU, and RSU platforms and do not depend on new hardware generations. Accordingly, short-term priorities include updates to the standards and certificate framework, hybrid credential support, handling fragmentation, and PQC field testbeds (RG1-RG4, RG6, RG10).

The medium term (approximately 5-10 years) covers gaps related to embedded optimization, SCMS architectural evolution, compression frameworks, and cross-stack interoperability. These items usually require updates to both software stacks and hardware platforms, and therefore, tend to follow multi-year platform refresh cycles with coordinated vendor support and field validation. This phase includes embedded PQC optimization, SCMS and provisioning redesign, improved compression and transport efficiency, and interoperability across DSRC, C-V2X, and ITS-G5 environments (RG5, RG7-RG9).

The long term (approximately 10-20 years) focuses on PQC-only infrastructures and advanced resilience mechanisms, such as trust-precedence enforcement and stronger protections against side-channel and fault-injection attacks. These goals typically require new hardware support, updated certification frameworks, and validated supply-chain components, and therefore cannot be achieved through software updates alone. For this reason, they align with hybrid-deployment security strengthening and physical-layer resilience gaps (RG11-RG13), and represent later-stage steps toward fully PQC-dominant ITS infrastructures.

To further clarify the practical nature of these research gaps, they can also be viewed by dominant constraint type. Some gaps are primarily technical and engineering-driven, where progress depends primarily on algorithmic optimization, software efficiency, and hardware acceleration support. These include embedded PQC optimization, artifact size reduction, and compression efficiency (RG5, RG7, RG8). Other gaps are standards- and coordination-driven, where the main challenge is achieving cross-organizational agreement and ecosystem alignment among relevant political and business stakeholders. These include certificate-format migration, hybrid credential/authentication frameworks, protocol updates, and cross-stack interoperability (RG1-RG4, RG9). A third group is constrained by economic and lifecycle factors, in which deployment depends on platform refresh cycles, certification costs, and infrastructure replacement timelines. This includes SCMS and provisioning redesign, as well as large-scale PQC testbed efforts (RG6, RG10). Finally, physical-layer gaps (RG11-RG13) are of a persistent-risk nature, where defenses must continuously evolve with the attacker's capability and should be treated as ongoing resilience requirements.

In addition, these constraint-based categories do not always align directly with urgency. Some coordination- and lifecycle-driven gaps are time-consuming yet remain high-priority because many downstream technical, certification, and deployment activities depend on them, particularly the standards and certificate framework gaps (RG1-RG4). In practice, updating standards is only one step toward deployment. Real-world adoption often takes significantly longer due to OEM inertia, liability concerns, recertification effort, integration cost, and supply-chain readiness. Therefore, standards approval should be viewed as a starting point, not as a signal that immediate field deployment will follow.

**Table 15**
**Phased roadmap for research and implementation of PQC in ITS**

| Phase | Potential Focus Areas | Associated Research Gaps | Potential Lead Stakeholders |
|---|---|---|---|
| Short-Term Priorities (Within ~5 years) | • Standards and certificate framework updates (e.g., IEEE 1609.2, ETSI TS 103 097),<br>• Hybrid certificate initialization, and<br>• Real-world performance testbeds | RG1, RG2, RG3, RG4, RG6, RG10 | • Standards bodies (e.g., IEEE, ETSI, NIST),<br>• Regulators (e.g., UNECE WP.29, ISO/SAE),<br>• OEMs, and<br>• Testbed consortia |
| Mid-Term Priorities (~5-10 years) | • Embedded PQC optimization,<br>• Message fragmentation handling,<br>• SCMS lifecycle extension, and<br>• Protocol-level interoperability (e.g., between C-V2X and ETSI ITS-G5) | RG5, RG7, RG8, RG9 | • ECU vendors,<br>• SCMS operators,<br>• Middleware developers, and<br>• Hardware designers |
| Long-Term Priorities (~10-20 years) | • SCA/FIA countermeasure integration,<br>• Trust precedence enforcement, and<br>• PQC-only infrastructure migration | RG11, RG12, RG13 | • Cybersecurity labs,<br>• OEMs,<br>• Certification bodies, and<br>• National cybersecurity agencies |

Notes- C-V2X: Cellular Vehicle-to-Everything; ECU: Electronic Control Unit; ETSI: European Telecommunications Standards Institute; FIA: Fault Injection Attack; IEEE: Institute of Electrical and Electronics Engineers; ISO: International Organization for Standardization; ITS: Intelligent Transport Systems; NIST: National Institute of Standards and Technology; OEM: Original Equipment Manufacturer; PQC: Post-Quantum Cryptography; SAE: Society of Automotive Engineers; SCA: Side-Channel Attack; SCMS: Security Credential Management System; UNECE: United Nations Economic Commission for Europe; WP.29: World Forum for Harmonization of Vehicle Regulations.

## 8. CONCLUSIONS

The advent of quantum computing poses a threat to the cryptographic foundations of ITS. Classical algorithms, such as ECC for secure communication between vehicles, infrastructure, and backend systems, are vulnerable to attacks from quantum computers. With the rapid advances in large-scale quantum computers, transitioning to PQC has become an operational necessity. This implementation-focused review examined the integration of PQC across in-vehicle, OEM, and V2X communication links from a deployment and engineering perspective, emphasizing implementation-level vulnerabilities, integration constraints, and system-level feasibility. The review also covered the vulnerabilities of classical cryptographic schemes used in ITS to quantum attacks, the current status of NIST PQC standardization as a migration pathway, compliance gaps in current vehicular communication and security standards, and emerging physical-layer threats, such as side-channel and fault-injection attacks, that challenge real-world deployment. Analysis of current standards reveals that standards, such as IEEE 1609.2, ETSI C-ITS, and ISO/SAE 21434, have not yet incorporated PQC algorithms or certificate structures. In-vehicle networks and OEM communications face bandwidth and hardware limitations, while V2X systems encounter latency and fragmentation challenges when transmitting large PQC payloads. Moreover, PQC implementations remain susceptible to side-channel and fault-injection attacks, especially in embedded automotive systems.

Across standards, communication, and physical-layer security domains, the survey of over 180 sources identifies thirteen structured research gaps (RG1-RG13). These include the lack of PQC-ready certificate formats in vehicular standards, unclear regulatory timelines, limited cross-certification between classical and PQC roots, and insufficient implementation-ready methods for message compression, pseudonym lifecycle management, and interoperability across DSRC, C-V2X, and ETSI ITS-G5 communication stacks. Implementation challenges persist due to unoptimized PQC algorithms for constrained ECUs and immature hybrid migration strategies. Even standardized schemes, such as ML-KEM (Kyber), ML-DSA (Dilithium), and SLH-DSA (SPHINCS+), impose significant computational and bandwidth overheads, while physical-layer vulnerabilities indicate the need for resilient hardware and countermeasures.

Future directions are likewise framed from an implementation and deployment perspective. Standards and certification efforts (RG1-RG4) are necessary to enable PQC-ready certificates and hybrid trust hierarchies. Implementation and migration (RG5-RG6) require hardware-optimized and energy-efficient PQC variants to support real-time operation. Protocol and interoperability research (RG7-RG9) can focus on network transmission mechanisms that reliably handle highly fragmented PQC payloads, standardized compression, and cross-stack compatibility between communication devices. Deployment and resilience priorities (RG10-RG13) involve real-world PQC testbeds and robust defenses against side-channel and fault-injection attacks.

Finally, the presented roadmap aligns short-, medium-, and long-term priorities with these implementation-focused research gaps, offering a staged path toward quantum-resilient ITS. Advances in cryptographic algorithms, communication and security protocols, and embedded protection mechanisms will be essential for strengthening the ITS ecosystem against quantum adversaries. Combined with sustained collaboration among regulators, industry, and academia, these advances can enable secure, scalable, and future-proof protection for ITS in the era of quantum computing.

## ACKNOWLEDGEMENT

This material is based upon the work supported by the National Center for Transportation Cybersecurity and Resiliency (TraCR) headquartered at Clemson University, Clemson, South Carolina, USA. Any opinions, findings, conclusions, and recommendations expressed in this material are those of the author(s) and do not necessarily reflect the views of TraCR, and the U.S. Government assumes no liability for the contents or use thereof.

## Declaration of Generative AI and AI-Assisted Technologies in the Manuscript Preparation Process

During the preparation of this work, the authors used ChatGPT (OpenAI) solely for grammar checking and limited paraphrasing to improve language clarity and readability. The tool was not used to generate scientific content, technical results, figures, tables, references, or any kind of analyses. All technical interpretations, evaluations, and conclusions were developed by the authors. After using this tool, the authors reviewed and edited the content as needed and take full responsibility for the content of the published article.